\documentclass{emulateapj}

\usepackage{graphicx}
\usepackage{color}
\usepackage{amsmath}
\usepackage{amssymb}
\usepackage{hyperref}
\usepackage{epstopdf}
\shorttitle{Fiber Fabry-Perot Interferometers as Stable Calibration Sources}
\shortauthors{Halverson et al.}
\slugcomment{Accepted in PASP}

\begin{document}

\title{Development of Fiber Fabry-Perot Interferometers as Stable Near-infrared Calibration Sources for High Resolution Spectrographs.}

\author{Samuel Halverson\altaffilmark{1,2,3}, 
Suvrath Mahadevan\altaffilmark{1,2,3}, 
Lawrence Ramsey\altaffilmark{1,2}, 
Fred Hearty\altaffilmark{4}, 
John Wilson\altaffilmark{4}, 
Jon Holtzman\altaffilmark{5}, 
Stephen Redman\altaffilmark{6}, 
Gillian Nave\altaffilmark{6}, 
David Nidever\altaffilmark{7}, 
Matt Nelson\altaffilmark{4}, 
Nick Venditti \altaffilmark{1},
Dmitry Bizyaev\altaffilmark{8}, 
Scott Fleming \altaffilmark{9}}

\altaffiltext{1}{Department of Astronomy \& Astrophysics, The Pennsylvania State University, University Park, PA 16802, USA}
\altaffiltext{2}{Center for Exoplanets \& Habitable Worlds, University Park, PA 16802, USA}
\altaffiltext{3}{Penn State Astrobiology Research Center, University Park, PA 16802, USA}
\altaffiltext{4}{Department of Astronomy, University of Virginia, P.O. Box 400325, Charlottesville, VA 22904, USA}
\altaffiltext{5}{Department of Astronomy, New Mexico State University, Box 30001, Las Cruces, NM 88003, USA}
\altaffiltext{6}{Atomic Physics Division, National Institute of Standards and Technology, Gaithersburg, MD 20899, USA}
\altaffiltext{7}{Department of Astronomy, University of Michigan, Ann Arbor, MI 48109, USA}
\altaffiltext{8}{Apache Point Observatory, P.O. Box 59, Sunspot, NM 88349, USA}
\altaffiltext{9}{Space Telescope Science Institute, 3700 San Martin Dr, Baltimore, MD 21211, USA}

\keywords{ \object{Astronomical Instrumentation} - \object{Extrasolar Planets}}

\begin{abstract}
We discuss the ongoing development of single-mode fiber Fabry-Perot (FFP) Interferometers as precise astro-photonic calibration sources for high precision radial velocity (RV) spectrographs. FFPs are simple, inexpensive, monolithic units that can yield a stable and repeatable output spectrum. An FFP is a unique alternative to a traditional etalon, as the interferometric cavity is made of single-mode fiber rather than an air-gap spacer. This design allows for excellent collimation, high spectral finesse, rigid mechanical stability, insensitivity to vibrations, and no need for vacuum operation.  The device we have tested is a commercially available product from Micron Optics\footnotemark[0]\footnotetext[0]{Certain commercial equipment, instruments, or materials are identified in this paper in order to specify the experimental procedure adequately. Such identification is not intended to imply recommendation or endorsement by the National Institute of Standards and Technology, nor is it intended to imply that the materials or equipment identified are necessarily the best available for the purpose.}. 
Our development path is targeted towards a calibration source for the Habitable-Zone Planet Finder (HPF), a near-infrared spectrograph designed to detect terrestrial-mass planets around low-mass stars, but this reference could also be used in many existing and planned fiber-fed spectrographs as we illustrate using the Apache Point Observatory Galactic Evolution Experiment (APOGEE) instrument. With precise temperature control of the fiber etalon, we achieve a thermal stability of 100 $\mu$K and associated velocity uncertainty of 22 cm s$^{-1}$. We achieve a precision of $\approx$2 m s$^{-1}$ in a single APOGEE fiber over 12 hours using this new photonic reference after removal of systematic correlations. This high precision (close to the expected photon-limited floor) is a testament to both the excellent intrinsic wavelength stability of the fiber interferometer and the stability of the APOGEE instrument design. Overall instrument velocity precision is 80 cm s$^{-1}$ over 12 hours when averaged over all 300 APOGEE fibers and after removal of known trends and pressure correlations, implying the fiber etalon is intrinsically stable to significantly higher precision. 
\end{abstract}

\section{Introduction}

Many hundreds of exoplanets have been discovered or confirmed using high precision Doppler radial velocity (RV) techniques. With high resolution spectrographs such as HIRES \citep{1994SPIE.2198..362V} on the Keck I telescope and HARPS \citep{2003Msngr.114...20M} on the ESO La Silla 3.6m telescope reaching long term precisions of 1 - 2 m s$^{-1}$ or better in the visible, detecting low mass planets around other stars has become a feasible endeavor.  Utilizing precise calibration and novel data reduction methods, a wealth of low-mass planets are being discovered using these instruments.
M-dwarfs are estimated to makeup the majority (over 70 percent) of stars in the Milky Way (RECONS\footnote{\url{http://www.chara.gsu.edu/RECONS/}}). These small stars have low masses, resulting in large Doppler reflex signals due to orbiting companions. Habitable-Zone (HZ) boundaries \citep{2013ApJ...765..131K} are also much closer to the host stars in M-dwarf systems than larger solar-type stars, making them ideal candidates for surveys designed to detect potentially habitable planets. Recently \cite{2013arXiv1302.1647D} used $Kepler$ data to estimate the frequency of terrestrial planets in the HZ of cool stars and arrive at a planetary occurrence rate of $0.15^{+0.13}_{-0.06}$ per star for Earth-size planets (0.5 - 1.4)$R_{\earth}$. \cite{2013ApJ...767L...8K} reanalyzed this data using new HZ boundaries and arrive at an even higher terrestrial planet frequency of $0.48^{+0.12}_{-0.24}$ per M-dwarf. These high occurrence rates imply the solar neighborhood is potentially teeming with terrestrial planets orbiting in the HZs of cool stars. This is an exciting region of exoplanet discovery space, and many dedicated instruments are being built specifically to observe these cool stars. Achieving 1 m s$^{-1}$ precision will allow for the robust detection of terrestrial-mass planets in the HZs of many nearby M-dwarfs systems. 

Instruments such as the Habitable-Zone Planet Finder (HPF, \cite{2012SPIE.8446E..1SM}) and the Calar Alto high-Resolution search for M dwarfs with Exoearths with Near-infrared and optical Echelle Spectrographs (CARMENES, \cite{2012SPIE.8446E..0RQ}) will be on the forefront of dedicated near-infrared (NIR) Doppler instruments, capable of achieving 1 - 3 m s$^{-1}$ measurement precision on low mass M-dwarfs. HPF consists of a stabilized, R $\approx$ 50 000 fiber-fed spectrograph enclosed in a large vacuum cryostat that is cooled to 180 K. The spectrograph simultaneously covers parts of the z (0.8 - 0.9 $\mu$m), Y (0.95 - 1.1 $\mu$m) and J (1.2 - 1.35 $\mu$m) NIR bands. For the next generation of stabilized NIR Doppler spectrographs to be able to achieve 1 m s$^{-1}$ precision or better on nearby stars, a highly stable calibration source must be used to accurately remove any instrumental drifts.

Detection of low mass planets requires stable, accurate instruments for confident detections. Simultaneous wavelength calibration of starlight, either by imprinting spectral features directly onto the stellar spectrum or by having a dedicated calibration fiber with a stable reference is required to achieve the highest measurement precisions. 
Wavelength calibrators such as molecular iodine cells and Thorium-Argon (Th-Ar) lamps have been used to great success in the visible as references against which Doppler shifts are measured, but can be limiting factors in the push for increasing measurement precision. The limited wavelength coverage and non-uniform spectral features of these references places limits on high precision measurements. This is especially true in the NIR, where no precise wavelength calibrators have been traditionally available.

Here we present test results of a commercially produced single-mode fiber (SMF) Fiber Fabry-Perot interferometer (FFP) as a precise calibration source in the H-band (1.5 - 1.7 $\mu$m). Tests with the Sloan Digital Sky Survey III (SDSS-III, \cite{2011AJ....142...72E}) APOGEE instrument demonstrate radial velocity precisions close to the expected photon-noise limit.  This device represents one of the only \lq{}astro-photonic\rq{} devices being developed for high precision RV measurements. FFPs, like conventional Fabry-Perot cavities, create interference patterns by combining light traversing different delays. The interference creates a rich spectrum of narrow lines, ideal for use as a precise spectrograph reference. An FFP can produce a high density of clean lines over a wide bandwidth, greatly expanding the wavelength regions over which precise RV measurements are possible. The physical nature of Fabry-Perot cavities does not, by itself, enable absolute wavelength calibration but does provide a stable grid of lines to track instrument drift.

\section{Summary of current precision Doppler wavelength references}
\label{sec:cals}

\subsection{Molecular Absorption Cells}

Simultaneous calibration using molecular Iodine (I$_2$) cells has achieved long-term RV precisions of 1$-3$ m s$^{-1}$\citep{1996PASP..108..500B, 2012ApJ...751L..16A} using the entire 500 - 620 nm bandwidth of the I$_2$ cell. Incident starlight is passed through a temperature controlled absorption cell which imprints the dense forest of I$_2$ molecular lines onto the stellar spectrum. This allows for precise correction of instrument point-spread function (PSF) changes and simultaneous wavelength calibration. The intrinsic I$_2$ spectrum must be measured to high accuracy in order to correctly model PSF variations and extract precise velocity measurements. No such absorption cell currently exists that covers major parts of the z/Y/J/H NIR bands, where the majority of the flux lies for a typical mid to late M-dwarf, though many development efforts are currently underway \citep{doi:10.1117/12.2023690}.

\cite{2010ApJ...713..410B} achieved $\approx$3 m s$^{-1}$ RV precision on bright M-dwarfs in the $K$ band using simultaneous calibration with a NH$_3$ cell on the CRIRES \citep{2004SPIE.5492.1218K} instrument on VLT. The wavelength coverage of this method is narrow, spanning roughly 36 nm. This also required the high resolving power (R = 100 000) of CRIRES and bright (K$<$8 mag) targets, but is an interesting wavelength region as $K$ band is densely populated with telluric absorption features from H$_2$O and CH$_4$.  

\cite{2010ApJ...723..684B} utilized this stable forest of telluric  H$_2$O and CH$_4$ bands near 2.3 $\mu$m to obtain RV precisions of $\approx$50 m s$^{-1}$ on ultracool dwarfs using the NIRSPEC instrument on the Keck II telescope. These telluric features are not static however, and require unique reduction methods and careful atmospheric monitoring to reach  $<$10 m s$^{-1}$ precision. 

\cite{2009ApJ...692.1590M} explored using a number of commercially available molecular gas cell references and concluded that a combination of H$^{13}$C$^{14}$N, $^{12}$C$_2$H$_2$, $^{12}$CO, and $^{13}$CO cells, illuminated by a continuum source, could provide a dense enough set of features for precise calibration in the H-band. \cite{2012SPIE.8446E..8GR} discuss the plans to improve the measurement precision of CRIRES using these cells.

\subsection{Emission Lamps}
Hollow cathode lamps (HCLs) are widely used in high-resolution spectroscopy to derive precise wavelength solutions. These lamps produce a dense set of emission features to calibrate against, though the density and relative strength  of these features can lead to a significant number of blends. Bright lines from the lighter fill gases of HCLs can be used for moderate precision wavelength calibration, but are a significant source of error in sensitive high-resolution instruments that require a high-precision calibration \citep{2006SPIE.6269E..23L}.

The popular Thorium-Argon HCL has been used to attain sub m s$^{-1}$ long term precision on stable, inactive Solar-type stars with HARPS \citep{2011arXiv1109.2497M}. The wealth of available atomic $^{232}$Th transitions  provide emission features spanning the UV to NIR regions, making the element ideal for use as a broadband wavelength reference. However Thorium does not contain sufficient bright lines beyond 1 $\mu$m to be the optimal wavelength calibration source in the NIR. Additionally, emission lines from the Ar fill gas in these lamps are bright in the NIR, and can saturate detector areas during typical exposure timescales. Ar lines are also susceptible to pressure shifts, making them highly sensitive to ambient conditions \citep{2006SPIE.6269E..23L}.

Precise calibration using Uranium-Neon (U-Ne) hollow-cathode lamps has yielded RV precisions of $<$10 m s$^{-1}$ using the Pathfinder testbed instrument \citep{2010SPIE.7735E.231R} on bright RV standard stars. $^{238}$U, like $^{232}$Th, is a heavy element with a long half-life. This represents one of the few emission lamps tested specifically for use in NIR Doppler instruments. A high density of lines is present in all commonly used NIR bands \citep{2011ApJS..195...24R, 2012ApJS..199....2R}, making the lamp ideal for use as a precise reference source. The U-Ne lamp in the NIR does not suffer from the bright Ar lines, though the high density of Uranium lines results in many blended features even at the high resolution of astronomical spectrographs. Both Th-Ar and U-Ne lamps are routinely used in the SDSS/APOGEE instrument.

\subsection{Laser Frequency Combs}
Laser Frequency Combs (LFC) have also been tested in astronomical applications with success in recent years. LFCs represent the pinnacle for current astronomical wavelength references, producing sharp emission features at frequencies traceable to a stable atomic transition. Combs used in astronomical applications comb are generally frequency stabilized by locking both the offset frequency and laser repetition rate to a global positioning system-stabilized atomic clock which can yield absolute precisions better than $10^{-10}$\citep{2012OExpr..20.6631Y}.

Despite the high cost and complexity in design, LFC development is a high priority in the spectroscopic community due to the desire to increase RV precision in order to detect Earth-like planets. The difficulty lies in creating a LFC that both spans the entire optical or NIR regions, and is stable over long intervals. LFCs have been shown to be stable to the cm s$^{-1}$ level \citep{2008Natur.452..610L}, but are not quite yet at the \lq{}turn-key\rq{} stage of use in astronomical instruments. While our group and other groups are actively pursuing this path \citep{2012OExpr..20.6631Y,2012OExpr..2013711P}, LFCs are currently expensive and are likely only a viable option for a limited number of high precision instruments on large telescopes.

\subsection{Fabry-Perot References}
Fabry-Perot interferometers (FPI) can yield high velocity precision with precise environmental control, though lack the innate absolute referencing of a frequency-locked LFC. To an astronomical spectrograph, the output spectrum of an FPI is quite similar to that of an LFC despite originating from very different physical properties. The intrinsic spectrum of the FPI is a picket fence of interferometric  Airy peaks filtered from a broadband continuum source, rather than a discrete set of emission lines generated from a series of laser pulses.

\cite{2012SPIE.8446E..8EW} tested custom air-gap FPIs for the HARPS and HARPS-N spectrographs, achieving 10 cm s$^{-1}$ stability over a nightly interval. In both instruments Zerodur was used as the cavity spacer material to minimize thermal sensitivity. Previous versions of the device were subject to large velocity drifts possibly due to varying collimation from the multi-mode fiber illumination \citep{2010SPIE.7735E.164W}. This was corrected by redesigning the interferometer housing to provide a more constant illumination incident on the cavity. Precise temperature and vacuum control are required to maintain the cavity length and density to the precision required. Night to night drift of the updated device is very low, averaging 10 cm s$^{-1}$ stability over  one night and 1 m s$^{-1}$ stability over 60 days \citep{2012SPIE.8446E..8EW}.
  
\subsection{Photon-Limited Precision Comparison of Sources}
\label{sec:photon_limit}
The ideal Doppler calibration spectrum is a picket fence of stable, sharp emission features.
Spectral regions with large flux gradients have the highest information content, and therefore contribute the highest precision. Continuum regions, areas with low line density, and areas with many blended lines do little to increase Doppler measurement precision significantly. A high density of uniform and fairly evenly separated lines will therefore give the highest velocity measurement precision \citep{2007MNRAS.380..839M}. Fabry-Perots and frequency combs have both of these attributes. For a given velocity shift, as measured by pixel $i$ at wavelength $\lambda(i)$ with associated noise $\sigma_{F}(i)$ in an extracted spectrum $F(i)$, the limiting velocity precision, $\sigma_v(i)$, is (as described in \cite{2001A&A...374..733B}):

\begin{equation}
\frac{\sigma_{v}(i)}{c} = \frac{\sigma_{F}(i)}{\lambda(i)[{\partial F(i)}/{\partial \lambda(i)}]}.
\label{eq:qfact}
\end{equation}

The total velocity precision, $\sigma_v$, from all pixels is then:

\begin{equation}
\sigma_v = \frac{1}{\sqrt{\sum_i \sigma_v(i)^{-2}}}.
\end{equation}

\begin{figure}
\begin{center}
\includegraphics[width=3.5in]{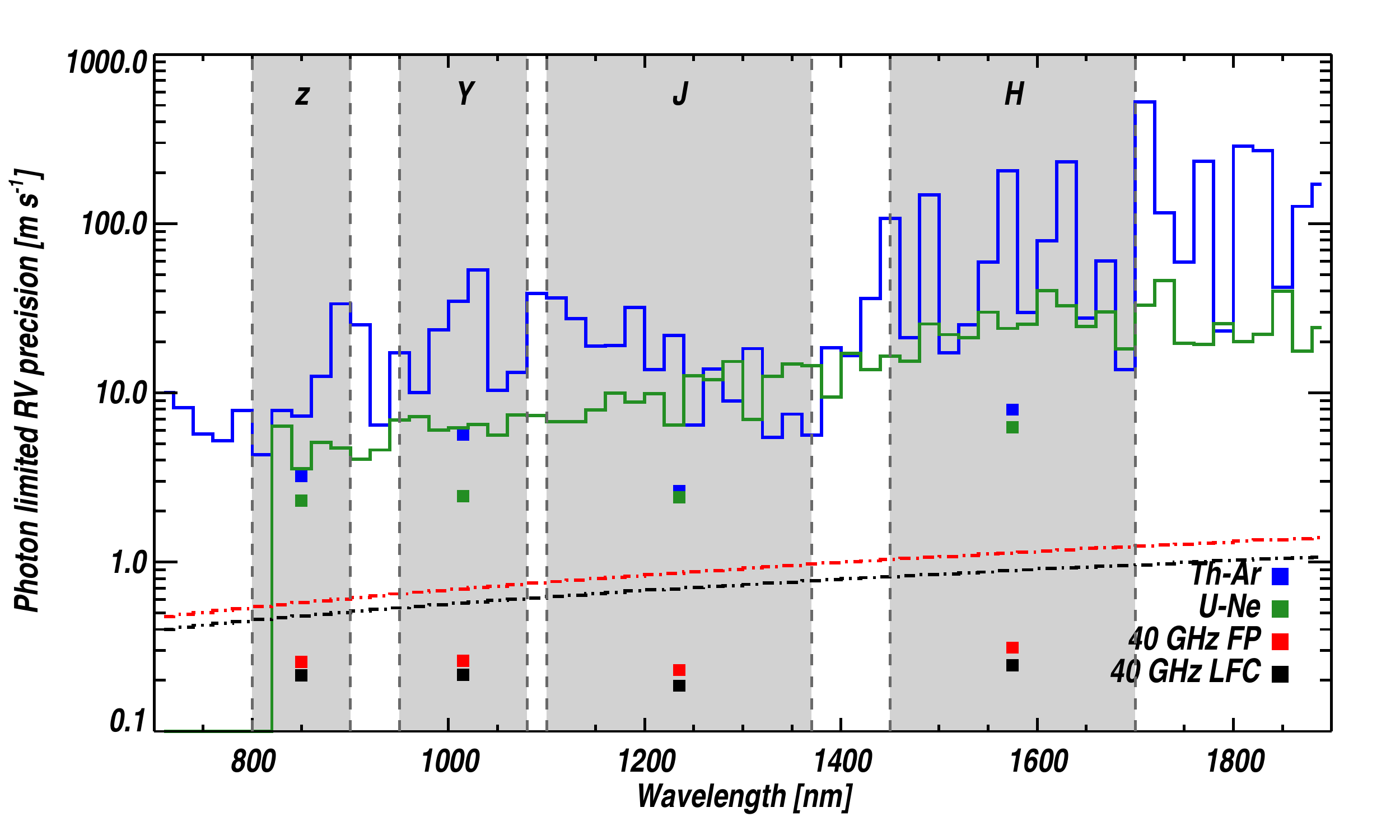}
\caption{Theoretical photon-limited velocity precision of reference sources discussed in the text using the methods from  \cite{2001A&A...374..733B}. Velocity errors were computed for 20 nm sections of model spectra assuming $R = $ 50 000 with three pixel sampling of the resolution element and maximum SNR of 200 in the extracted spectrum. The peak signals for the simulated LFC and FP spectra are assumed to be uniform for all lines. The squares are the total velocity precision in each NIR  band. The LFC and FP references give significant precision advantages over classical emission lamps.}
\label{fig:photon_rv}
\end{center}
\end{figure}

Figure~\ref{fig:photon_rv}  shows the expected photon-limited velocity measurement precision for several calibration sources discussed in the text assuming a noiseless detector, $R\approx$50 000 instrument resolving power, maximum signal-to-noise of 200 per pixel in extracted spectrum, and three-pixel sampling of the resolution element. The LFC and FP references clearly enable higher measurement precision than typical atomic emission lamps.
 Atomic lamp spectra are generated from archival FTS line lists \citep{2011ApJS..195...24R,0067-0049-178-2-374}. The fill gas elements (Argon, Neon) were not included in the analysis as they are generally excluded for precise calibration work. Emission lines from the fill gases also tend to be much brighter, exacerbating extraction issues.

\section{Fiber Fabry Perot Interferometer}
\label{sec:hardware}
\begin{figure}
\begin{center}
\includegraphics[width=3.4in]{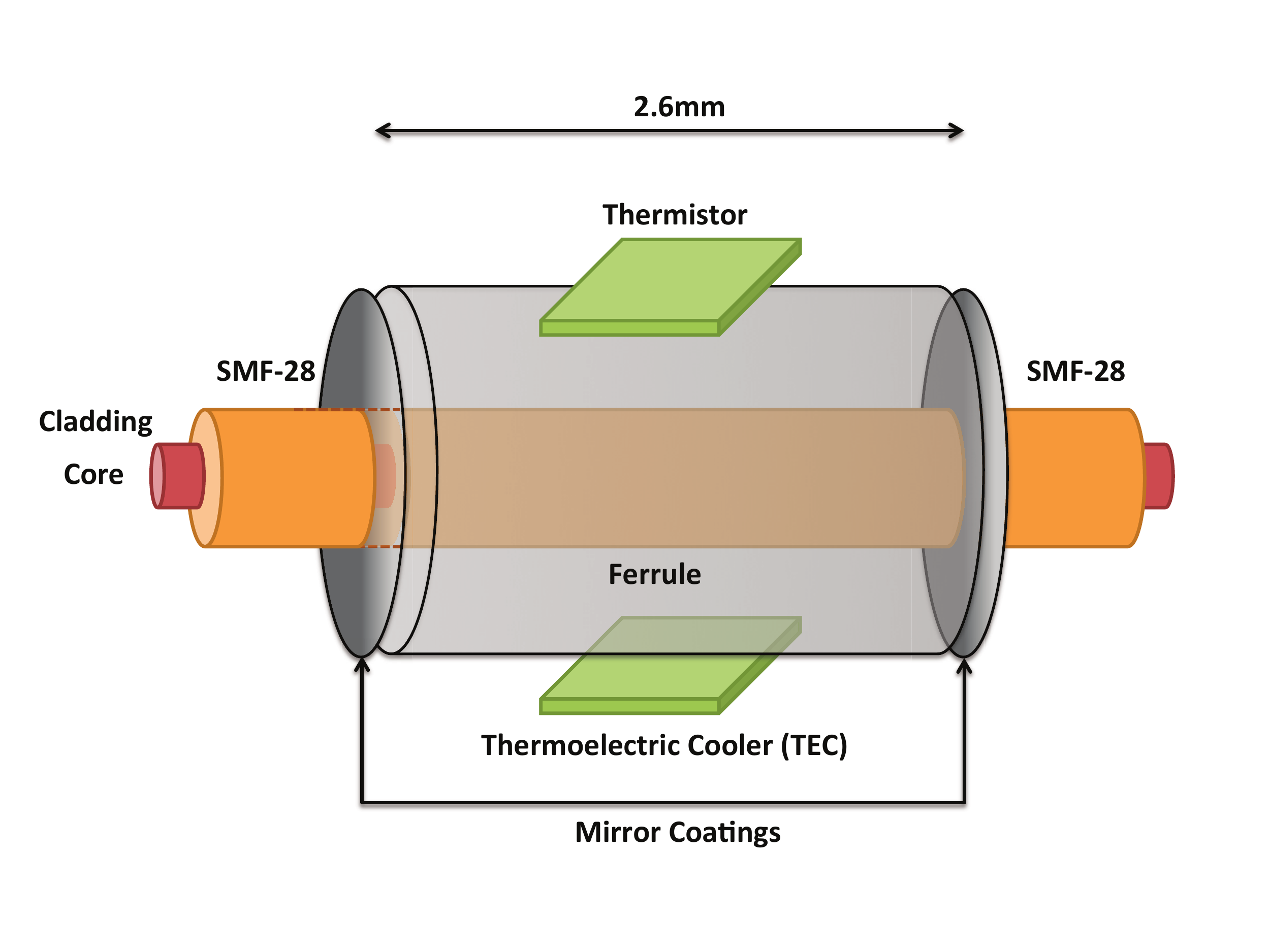}
\includegraphics[width=3.4in]{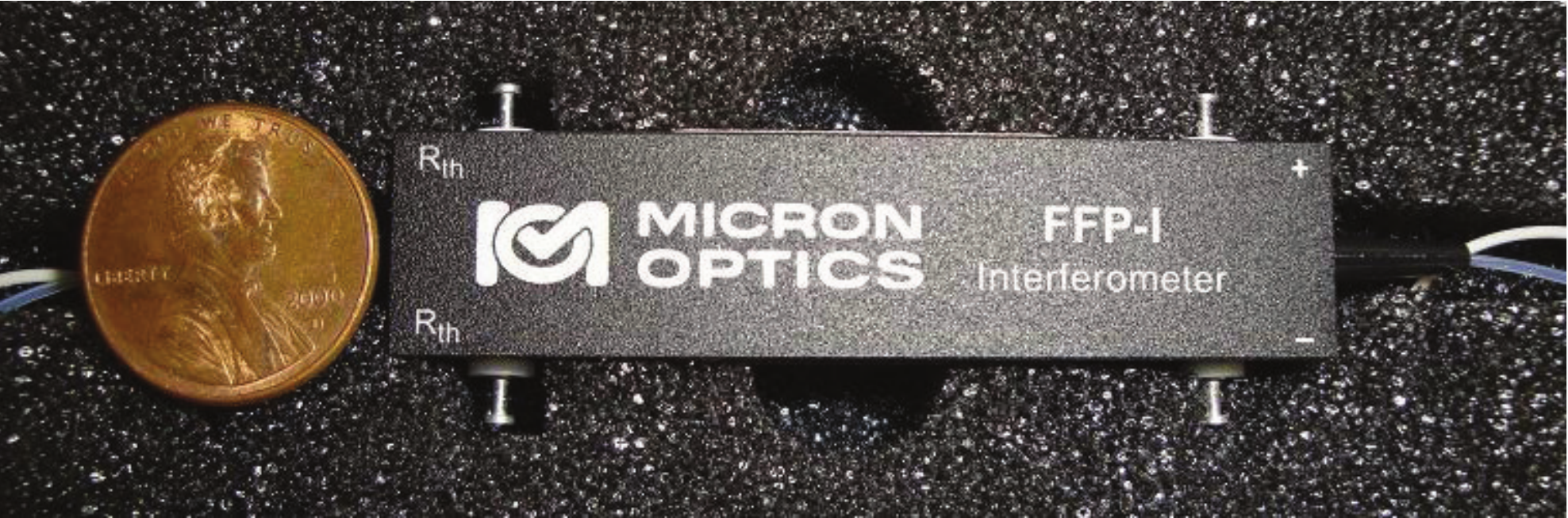}
\caption{Top: Internal schematic of FFP device. The device consists of SMF-28 fiber and custom coated dielectric mirrors. A thermistor sensor and thermo-electric cooler are epoxied directly to the central cavity section. Bottom: Picture of Micron-Optics H-band FFP. The pins on either side of the casing are leads for the temperature control system.}
\label{fig:ffp_schem}
\end{center}
\end{figure}

A full description of the FFP support hardware is described in \cite{doi:10.1117/12.925716}. A brief summary of the major components is presented here. The FFP used here is a monolithic commercially available unit manufactured by Micron Optics\footnotemark[0,]\footnote{Micron Optics FFPI$^{\mathrm{TM}}$}. The device consists of an input fiber epoxied to a central SMF interferometer cavity between two highly reflective dielectric mirrors, and an output SMF (see Figure~\ref{fig:ffp_schem}). Corning\footnotemark[0] SMF-28 fiber is used as the fiber waveguide. Input and output fibers are terminated with standard 2.5 mm FC connectors. The fiber cavity is epoxied into a solid ferrule for added rigidity.

The single-mode fiber waveguide ensures a stable beam propagates through the interferometer, enabling high finesse values and high stability. The central ferrule that contains the cavity fiber is small enough to be temperature controlled with bench-top controllers at high precision, as we demonstrate in the following section. 
The manufactured cavity size of 2.6 mm corresponds to an approximate frequency spacing of 40 GHz, equivalent to approximately 0.34 nm wavelength spacing (roughly 4.5 APOGEE resolution elements) in the H-band. Cavity spacing is chosen to minimize interline continuum while maximizing spectral information content. The selected dielectric mirror coatings have finesse of 36 across the H-band (1.5 - 1.7 $\mu$m) and very low losses. The finesse of the mirrors was chosen to yield under-resolved features at the APOGEE instrument resolution and maintain sufficient peak flux in the convolved spectrum.

\begin{table}
\begin{center}
\caption{Optical and thermal properties of SMF-28 fiber at 1550 nm, 25$^o$ C}
\begin{tabular}{cc}
\hline
\hline
Parameter & Value \\
\hline
\\
Mode Field Diameter [$\mu$m]		&	10.4 	\\
Single-mode Wavelength Range [nm] & 1260 - 1640 	\\
Effective Refractive Index, $n$	&	1.468 		\\
$\dfrac{\mathrm{d}n}{\mathrm{d}T}$ (K$^{-1}$)		&	$1.06\times10^{-5}$	 \\
$\dfrac{1}{L}\dfrac{\mathrm{d}L}{\mathrm{d}T}$	 (K$^{-1}$)&	$5.61\times10^{-7}$	 \\
\\
\hline
\label{tab:ffp_params}
\end{tabular}
\end{center}
\end{table}

\begin{figure}
\begin{center}
\includegraphics[width=3.5in]{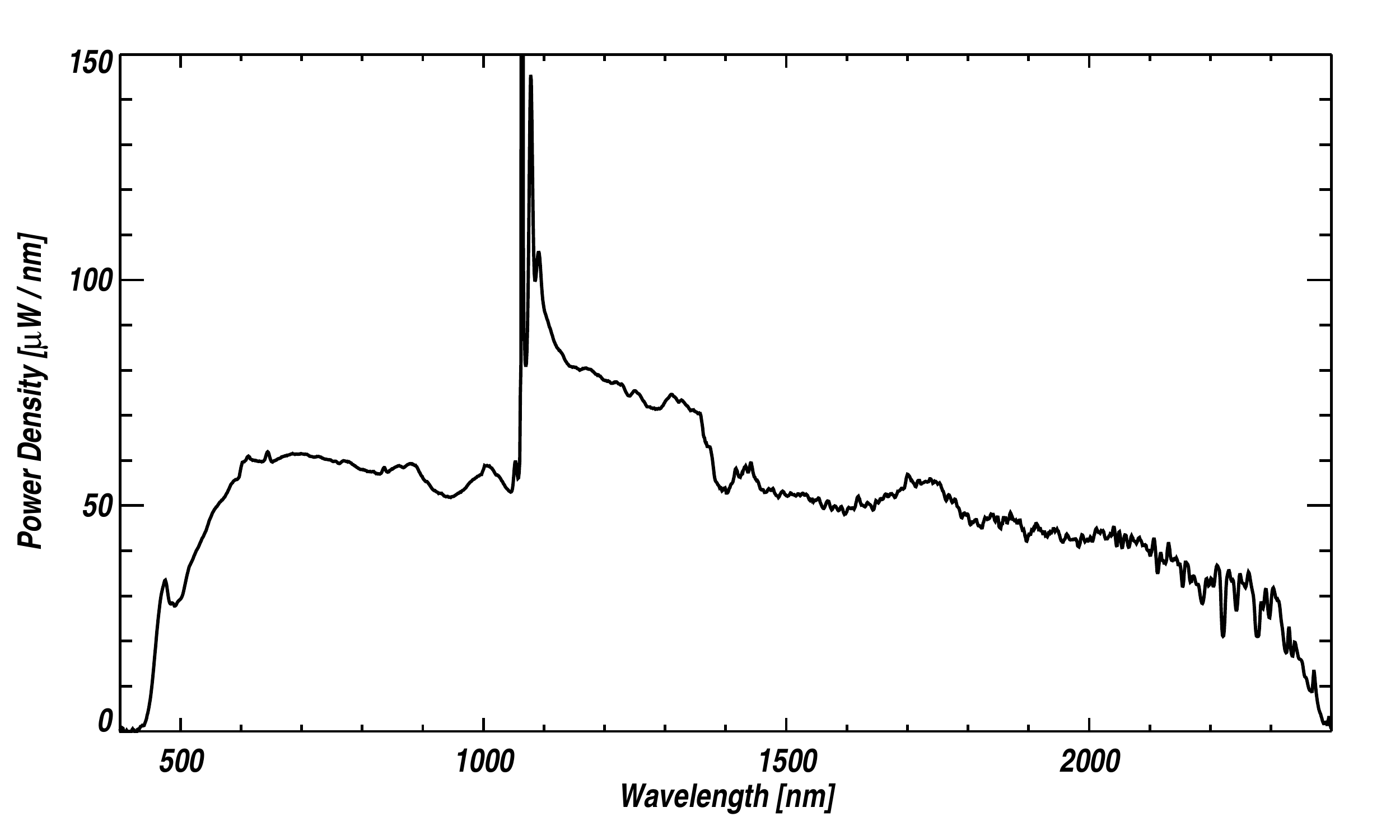}
\caption{Typical spectrum of SuperK Compact supercontinuum light source.}
\label{fig:superk_spec}
\end{center}
\end{figure}

As the device uses a single-mode waveguide as the cavity material, a supercontinuum source or fiber-coupled superluminescent diode is necessary for illumination.  Typical broadband lamps and light-emitting diode sources do not couple sufficient flux into SMFs to be used to illuminate the FFP. An NKT\footnotemark[0] supercontinuum light source is used as the illumination source, as it efficiently couples large amounts (100 mW integrated) of light directly into a SMF. Figure~\ref{fig:superk_spec} shows the typical power distribution for the supercontinuum source. Roughly 2 mW of integrated power is present in the H-band. A NIR dichroic is inserted prior to the FFP to filter out light below 900 nm. The resulting power distribution is coupled efficiently into an SMF-28 delivery fiber and coupled into the FFP input fiber using a mating sleeve.

\subsection{Temperature Stability}
The temperature sensitivity of the interferometer places a fundamental limit on the spectral stability. Typical thermal properties of SMF-28 fiber are listed in Table~\ref{tab:ffp_params}. The wavelength stability is dominated by refractive index variation with temperature, $\mathrm{dn}/\mathrm{dT}$. This requires that the cavity be stable to $<$500 $\mu$K in order to reach a velocity stability of 1 m s$^{-1}$.

\begin{figure}
\begin{center}
\includegraphics[width=3.5in]{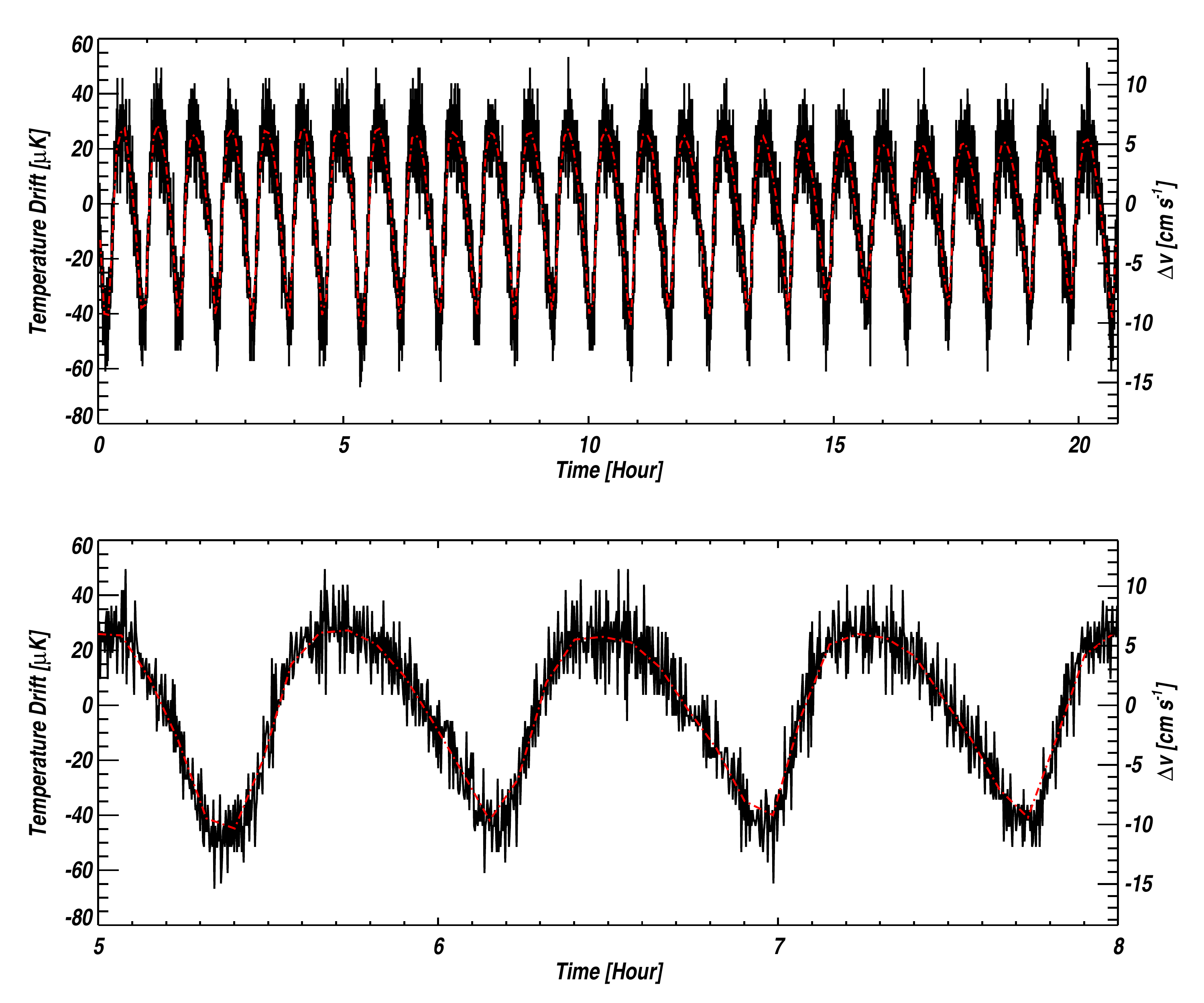}
\caption{Measured stability of FFP temperature control system in laboratory conditions, with approximate velocity drift (right vertical axis). The cavity is set to 298 K, roughly five degrees above ambient room temperature. The peak-to-peak temperature variation of the cavity is 100 $\mu$K over a 24 hour interval, corresponding to an expected velocity stability of 22 cm s$^{-1}$.}
\label{fig:ffp_temp}
\end{center}
\end{figure}

The FFP has a 10 k$\Omega$ thermistor and 0.5 Amp thermo-electric cooler (TEC) in direct contact with the cavity (see Figure~\ref{fig:ffp_schem}). The TEC and thermistor are used together in a proportional-integral-derivative (PID)  feedback loop to precisely maintain a specified cavity temperature. An off-the-shelf Stanford Research Systems temperature controller is used to read the on-board thermistor and control the TEC. The controller allows for sub-milliKelvin temperature control for small devices with good thermal contact and fast response times. The electronics noise floor of our controller is estimated by measuring the apparent temperature of a fixed resistance source (Vishay\footnotemark[0] 10 k$\Omega$ high precision resistor). The RMS measurement noise of the reference resistor temperature was 50 $\mu$K. The measured temperature stability of the FFP is 100 $\mu$K (see Figure~\ref{fig:ffp_temp}), close to the electronics noise floor of the temperature controller. This corresponds to an approximate velocity stability of 22 cm s$^{-1}$ based on the thermal properties of the SMF-28 fiber used in the etalon. This excellent control precision is a result of both a small cavity and good thermal coupling with the thermistor and TEC unit. 

We show in the next section that the device demonstrates 50 - 80 cm s$^{-1}$ stability over a 12 hour period, close to the expected measurement noise floor when considering temperature control precision and photon-noise alone.
\section{Integration and Performance on APOGEE}
\label{sec:APOGEE}
The APOGEE instrument, as part of the SDSS-III survey, is surveying 100 000 red stars spanning a large range of galactic latitudes to study the chemical and kinematical history of all Milky Way populations.  The instrument is a recently commissioned multi-object, 300 fiber R$\approx$22,500 H-band fiber-fed spectrograph \citep{2012SPIE.8446E..0HW} at the 2.5 m Sloan telescope \citep{2006AJ....131.2332G} at Apache Point Observatory. APOGEE is a cold (80 K), stabilized instrument enclosed in a vacuum cryostat. The RV precision requirement goal for the instrument is $<$300 m s$^{-1}$ \citep{2011AJ....142...72E} for the primary survey. 

\begin{figure*}
\begin{center}
\includegraphics[width=5in]{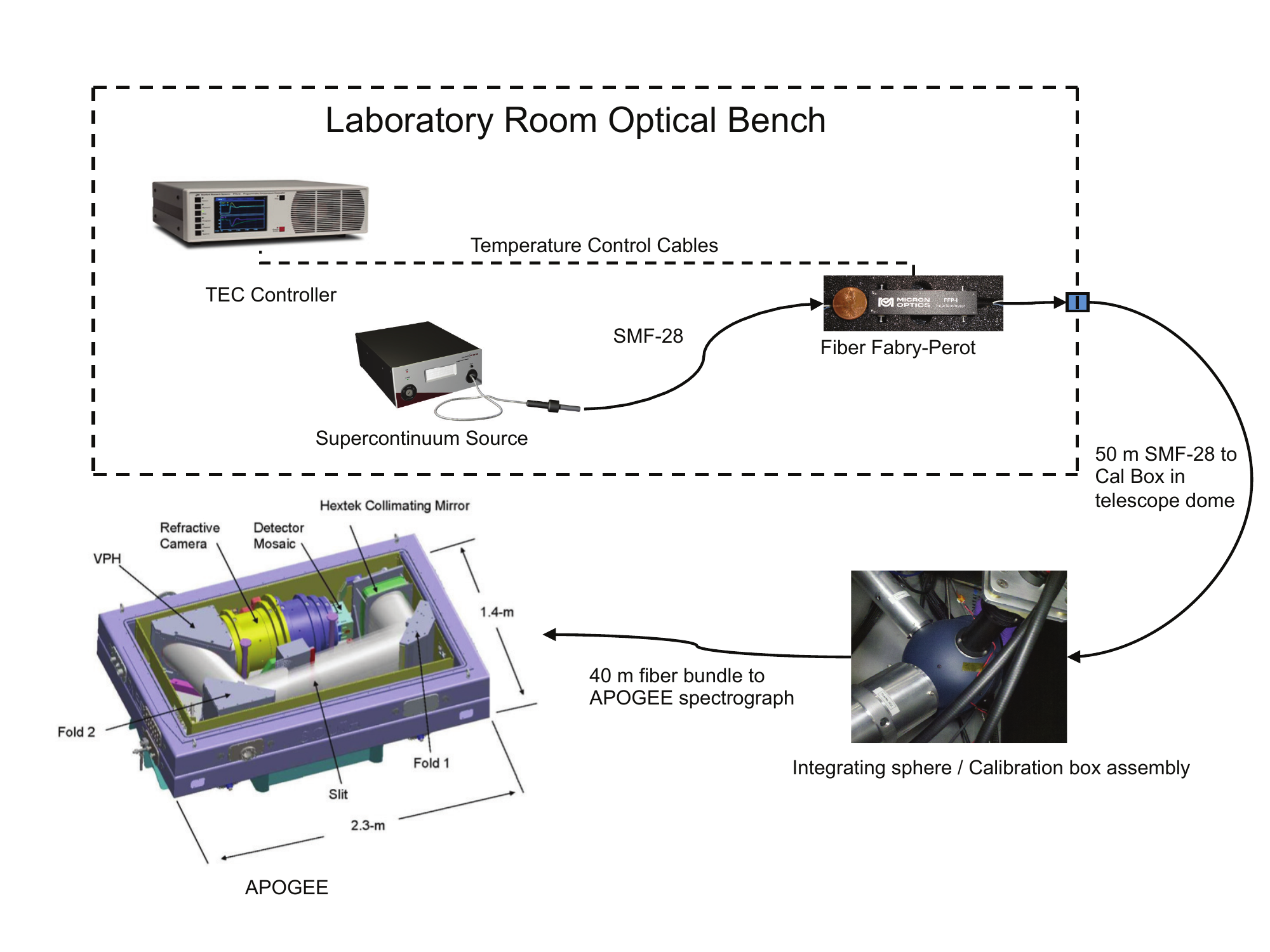}
\caption{On-mountain setup of FFP with APOGEE spectrograph. All FFP components were placed on a small optical breadboard in a temperature controlled room at APO. Light from the SuperK supercontinuum source is coupled to the FFP through a SMF-28 patch cable. The NIR fiber is used as the input fiber to the FFP. The output fiber of the FFP is coupled to 50 m of SMF-28 fiber and connected to the integrating sphere that houses the standard APOGEE calibration sources. Figure adapted from \cite{doi:10.1117/12.925716}}.
\label{fig:ffp_apogee_setup}
\end{center}
\end{figure*}

Such an instrument is a prime candidate for the addition of a stable, precise wavelength reference source to augment its existing calibration toolbox of traditional Th-Ar and U-Ne emission lamps and enable high RV measurement precision on hundreds of targets per night. Figure~\ref{fig:ffp_apogee_setup} shows the integration of the H-band FFP with the APOGEE instrument. At these resolutions the spectrum of a high finesse etalon will be not easily distinguishable from that of an LFC.

\begin{figure*}
\begin{center}
\includegraphics[width=6in]{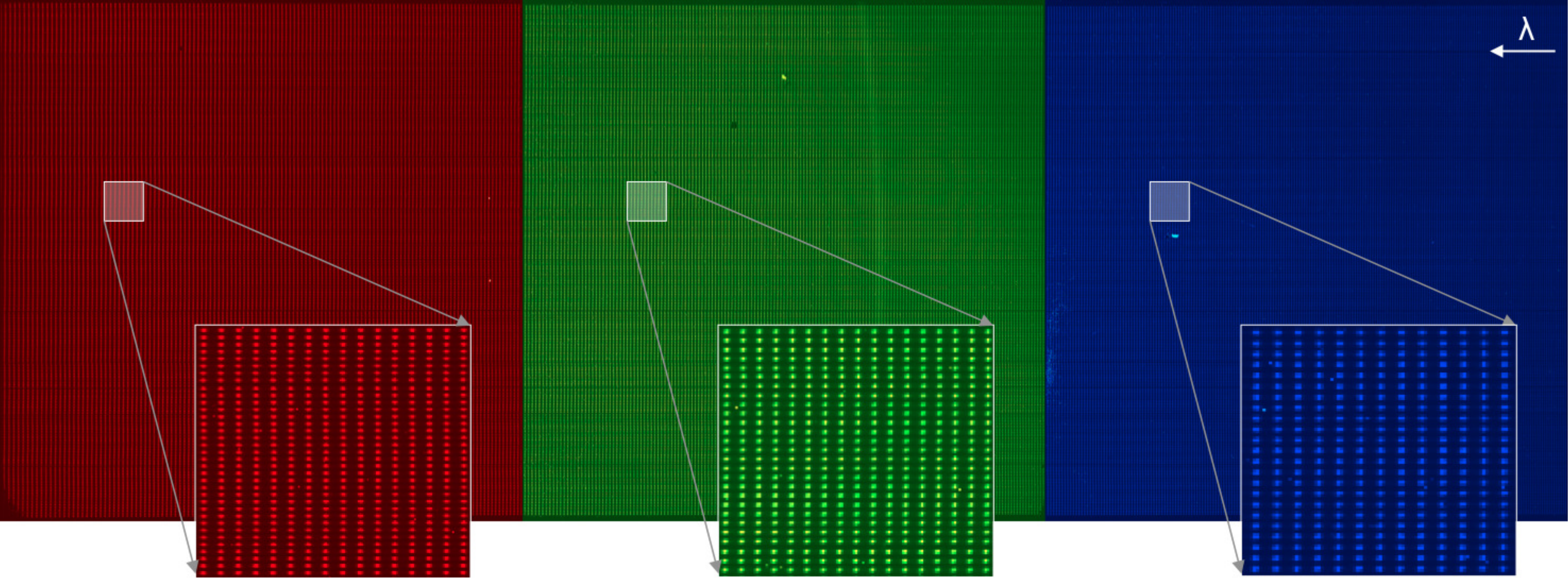}
\caption{First light frame of fiber Fabry-Perot on the APOGEE spectrograph, consisting of three detectors with wavelength coverage from 1514 nm on the right to 1696 nm on the left. The 300 individual fibers from the telescope are arranged vertically, yielding over 120 000 interferometric features across the detector mosaic. The red, green and blue colors of the chips correspond to the colors in Figures~\ref{fig:fiber_track} - \ref{fig:apogee_ffp_rv_final}.}
\label{fig:mosaic}
\end{center}
\end{figure*}

The FFP calibration system was tested on the APOGEE spectrograph at the 2.5 m Sloan telescope in Spring 2012. Light from the FFP was injected into the integrating sphere that houses the standard APOGEE calibration suite through a 50 m SMF patch cable. The FFP was kept in an adjacent building with moderate temperature control. Light from the integrating sphere and calibration system is fed into the spectrograph through a fiber bundle coupled to the 300 primary APOGEE science fibers.

The first light frame of the FFP on the three-detector APOGEE mosaic is shown in Figure~\ref{fig:mosaic}. While the intrinsic FFP features are narrow compared to the spectrograph response, their width is not negligible. The observed line-widths in the APOGEE focal plane are $\approx$6 \% wider than a typical resolution element. This is expected based on the specified finesse and FSR values of the interferometer.
\subsection{Observations on APOGEE}
 
 Several hundred FFP exposures were taken over the two-night observing period. The default APOGEE observation sequence is an ABBA ABBA dithering of the detector mosaic between two positions offset by 0.5 pixels. We chose not to use this observation mode as it may introduce added instrument instabilities due to detector motion, and instead opted for fixed-detector observations. Data for each of the three Hawaii-2RG detectors is extracted independently for each fiber. Instrument drifts are calculated separately for each detector to better separate any differences in systematic drifts. Exposure times when using the FFP were typically 6 minutes, resulting in a peak flux of roughly 40 000 counts in the brightest regions for a single exposure. This flux level was chosen to achieve high signal to noise in relatively short exposures while avoiding nonlinear detector effects\footnote{The typical full well depth for the detectors is roughly 100 000 ADU, corresponding to 50 000 counts. }.

\begin{figure}
\begin{center}
\includegraphics[width=3.5in]{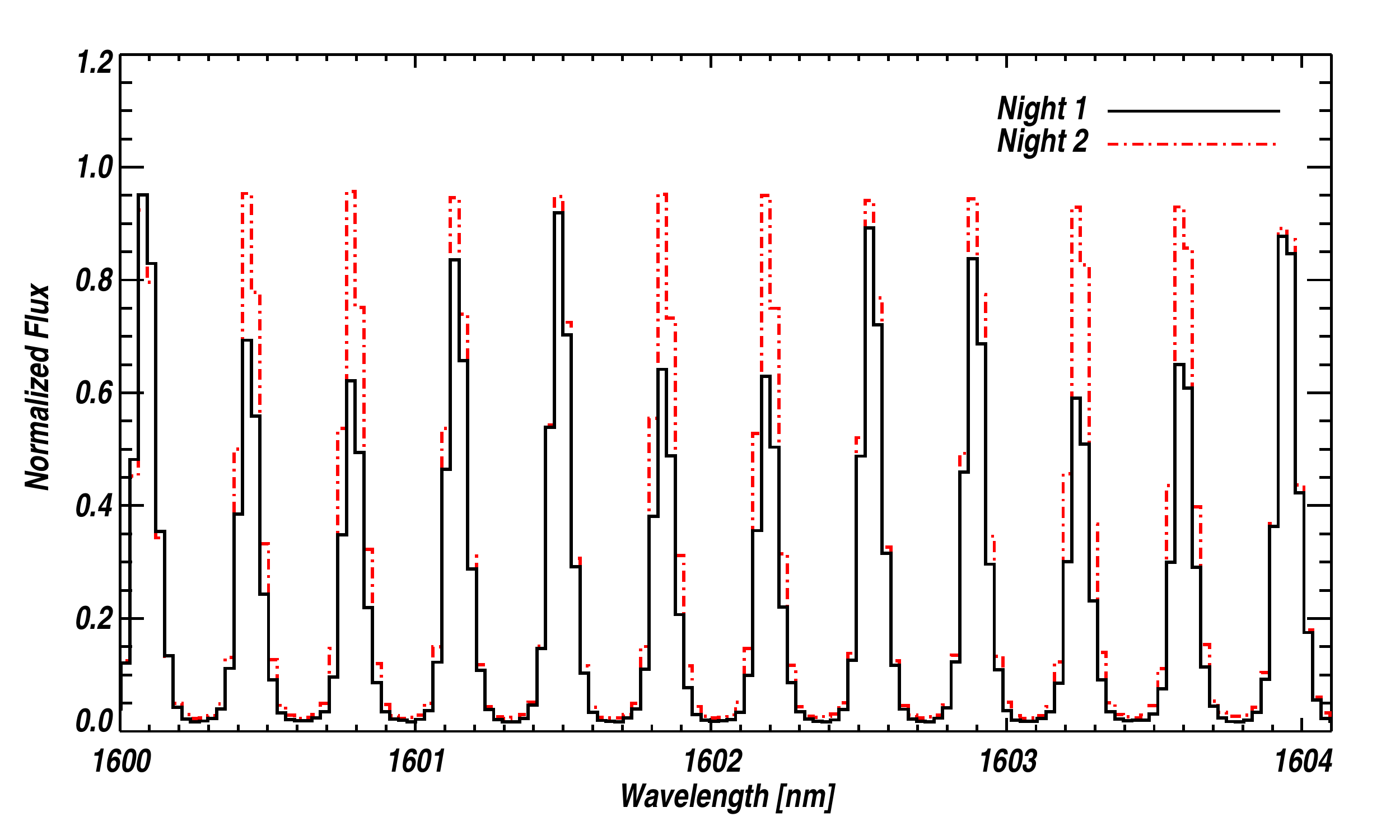}
\caption{Example FFP spectra on APOGEE for the same fiber on both nights of observations. The intensity modulation in the spectrum from the first night (black) is due to fringing from the ND filter placed at the output of the supercontinuum source.}
\label{fig:ap_spec_example}
\end{center}
\end{figure}

Our initial optical setup included a neutral density filter at the output of the supercontinuum source to reduce the risk of severely saturating the detectors. This filter introduced noticeable spectral fringing in data taken during the first night (see Figure~\ref{fig:ap_spec_example}). We removed the filter prior to the second night and decreased the output intensity of the supercontinuum source to a comparable level by slightly misaligning the fiber coupling into the FFP.
 
 \subsection{Calculating Instrument Drift}

\begin{figure}
\begin{center}
\includegraphics[width=3.5in]{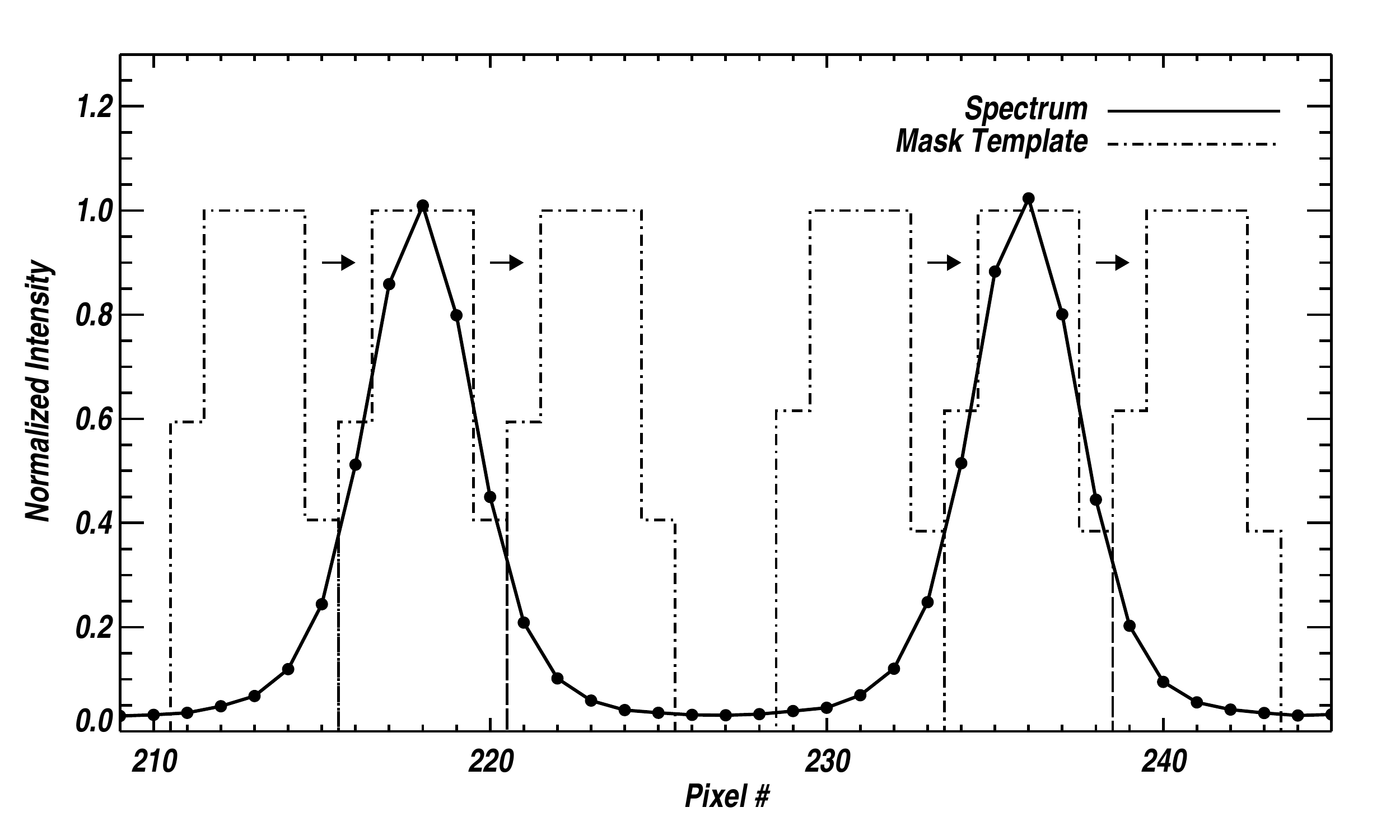}
\includegraphics[width=3.5in]{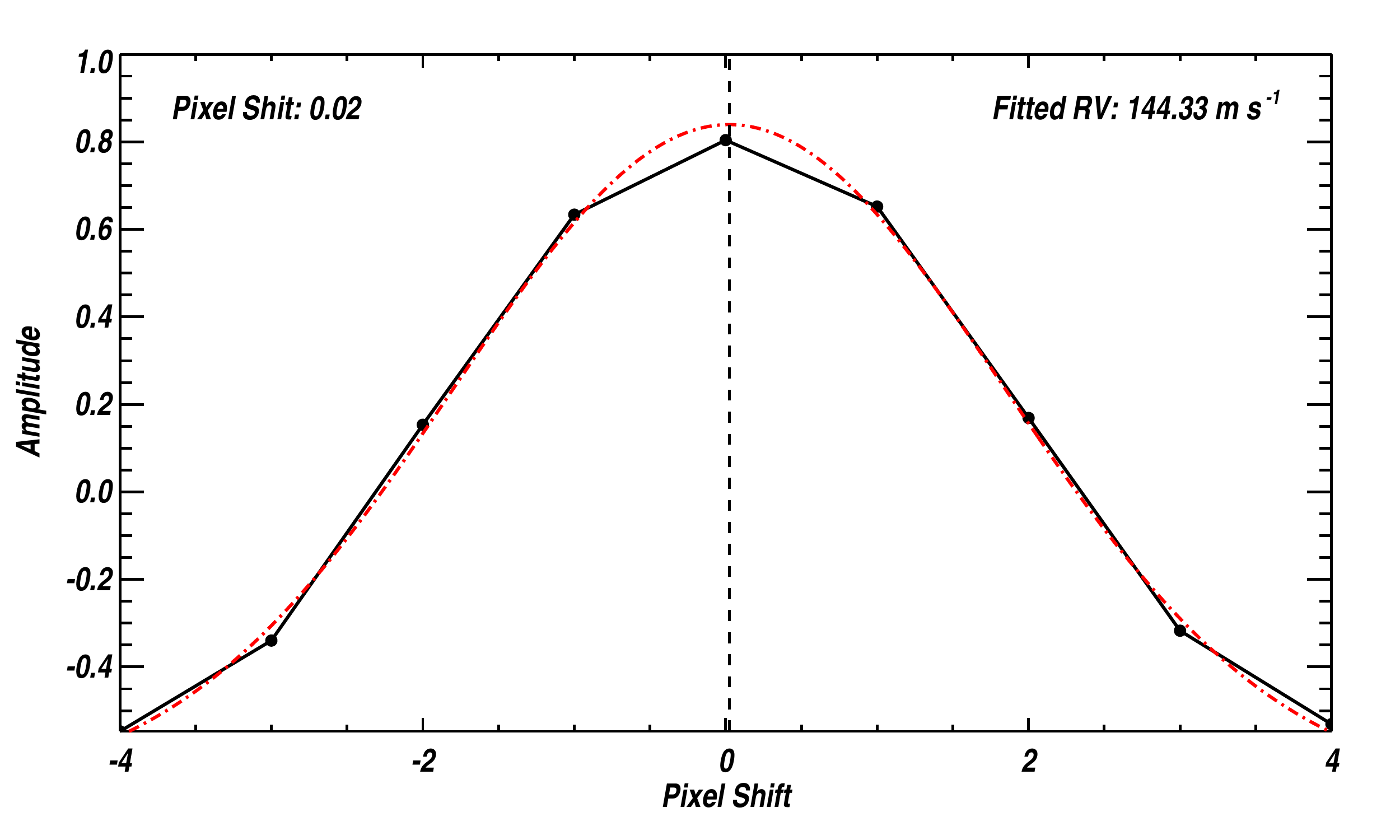}
\caption{Illustration of pixel mask cross-correlation technique used to derive instrument drifts from FFP data. Top: Each fiber spectrum is cross-correlated with a mask template derived from an averaged spectrum. Bottom: The cross-correlation function is sampled at different mask template pixel shifts and fit with a Gaussian function to determine the pixel offset.}
\label{fig:mask_rv}
\end{center}
\end{figure}

To precisely characterize instrument drift, we generate a template spectral mask using all FFP lines for each fiber from an averaged spectrum for each detector on the APOGEE mosaic. Mask lines are centered on measured pixel centers of FFP lines in the averaged spectrum, with a width of five pixels (Figure~\ref{fig:mask_rv}). All template lines are assigned equal mask weights, as the intrinsic signal of the data determines the cross-correlation weight. Radial velocities are derived by cross correlating the extracted FFP spectrum with the corresponding template mask for that particular fiber. The cross correlation function is then fit with a simple Gaussian function to derive the pixel offset, which is then converted to a velocity shift using the average wavelength dispersion across the detector. This technique is an adaptation of the CCF mask technique discussed in \cite{1996A&AS..119..373B} (and also described in \cite{2013arXiv1312.5471C}).

\begin{figure*}
\begin{center}
\includegraphics[width=3.5in]{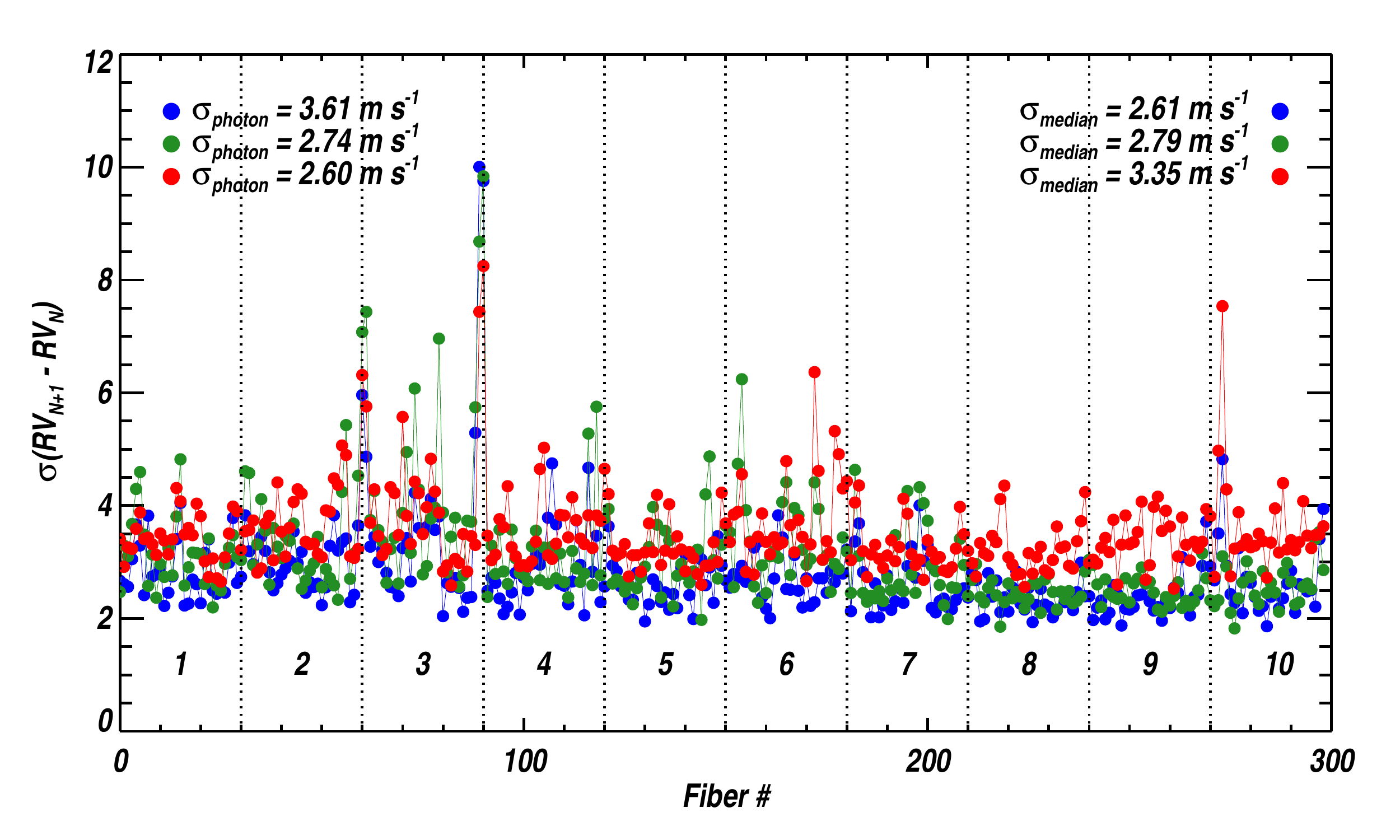}
\includegraphics[width=3.5in]{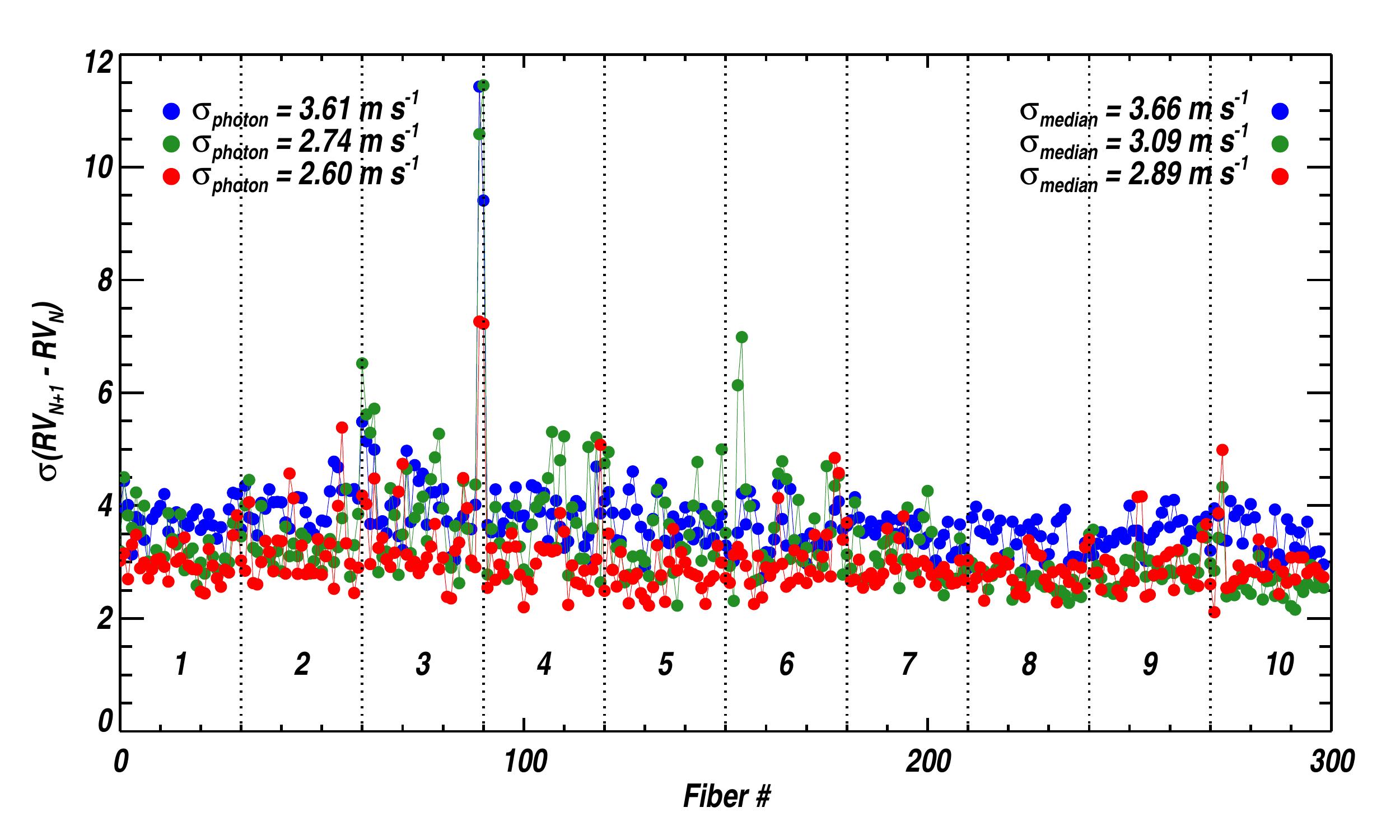}
\caption{Measured scatter when subtracting adjacent fiber drifts for night one (top) and night two (bottom). For both nights of observations, neighboring fibers track each other at close to the photon-limited measurement precision, $\sigma_\mathrm{photon}$. The vertical dashed lines indicate the location of each of the ten fiber mounting blocks comprising the pseudo-slit. Fibers with the largest deviations (e.g. 60 and 90) are on the edges of their respective v-groove blocks on the pseudo-slit. The different colors correspond to different detectors in the APOGEE mosaic.}
\label{fig:fiber_track}
\end{center}
\end{figure*}

\begin{figure}
\begin{center}
\includegraphics[width=2.4in,angle=90]{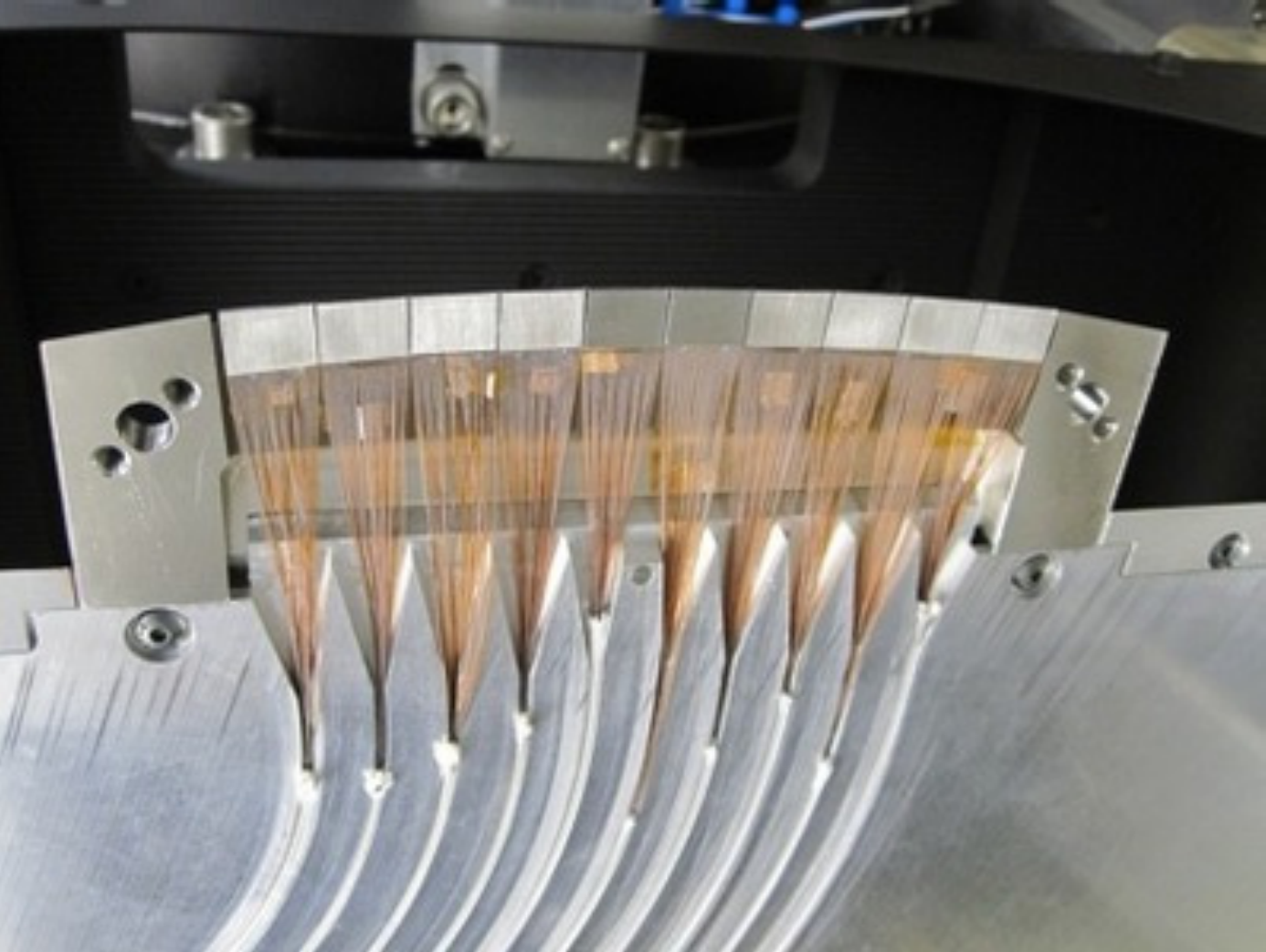}
\includegraphics[width=3.4in]{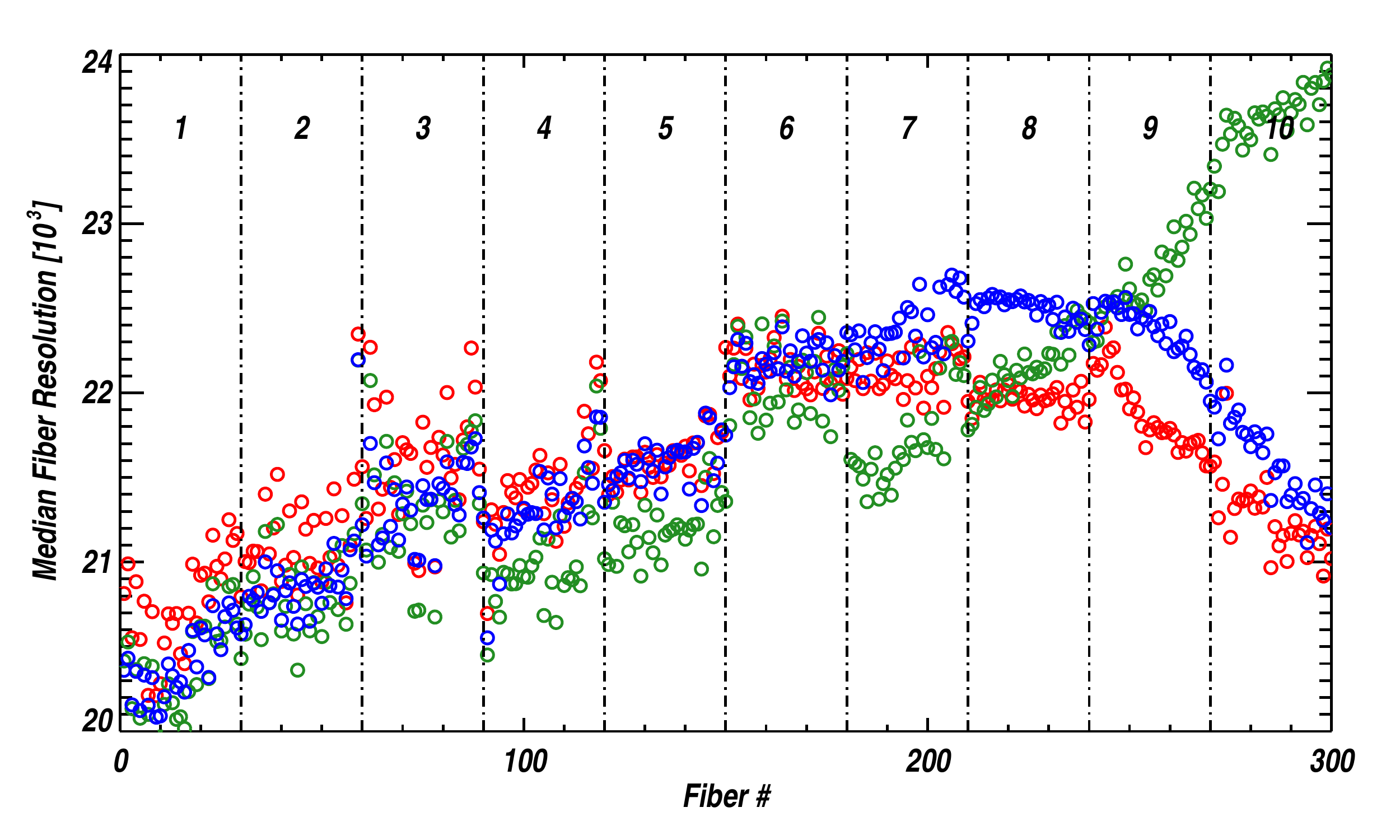}
\caption{Top: Image of APOGEE fiber pseudo-slit. The 10 fiber v-groove blocks comprising the pseudo-slit hold the 300 APOGEE science fibers. Image credit: SDSS-III Collaboration. Bottom: Median resolving power ($\lambda/\Delta\lambda$) of all 300 fibers for each of the three APOGEE detectors as measured using the FFP lines. Note that these values are approximately 6 \% lower than the intrinsic instrument resolution due to the finite width of the FFP features. Vertical dashed lines show the location of the 10 fiber v-grooves.}
\label{fig:vgroove_block}
\end{center}
\end{figure}

We find the median successive RV difference to be $\approx$3 m s$^{-1}$ between neighboring fibers. This high precision is close to the expected photon-noise limited RV precision for a single exposure, as calculated using the methods described in Section~\ref{sec:photon_limit} . Figure~\ref{fig:fiber_track} shows the relative tracking precision between two adjacent fibers for each night of data. The majority of fiber pairs track each other to close to the photon-limited precision, though certain fiber pairs deviate significantly. These large deviations are attributed to PSF variations between different fibers. The 300 APOGEE fibers are arranged in a single fiber pseudo-slit (see Figure~\ref{fig:vgroove_block}) consisting of 10 individual fiber v-groove blocks, each holding 30 fibers. Figure~\ref{fig:vgroove_block} shows the measured spectral resolving power, in units of $\lambda/\Delta\lambda$, for all 300 fibers as measured using the FFP lines. Towards the edges of the fiber blocks the scatter in spectral resolution increases significantly, leading to a large scatter in relative drifts between adjacent fibers. Adjacent fibers in the same mounting grooves yield the best tracking precisions.

\subsection{Drift Due to Liquid-Nitrogen Fill and Other Systematics}

\begin{figure}
\begin{center}
\includegraphics[width=3.5in]{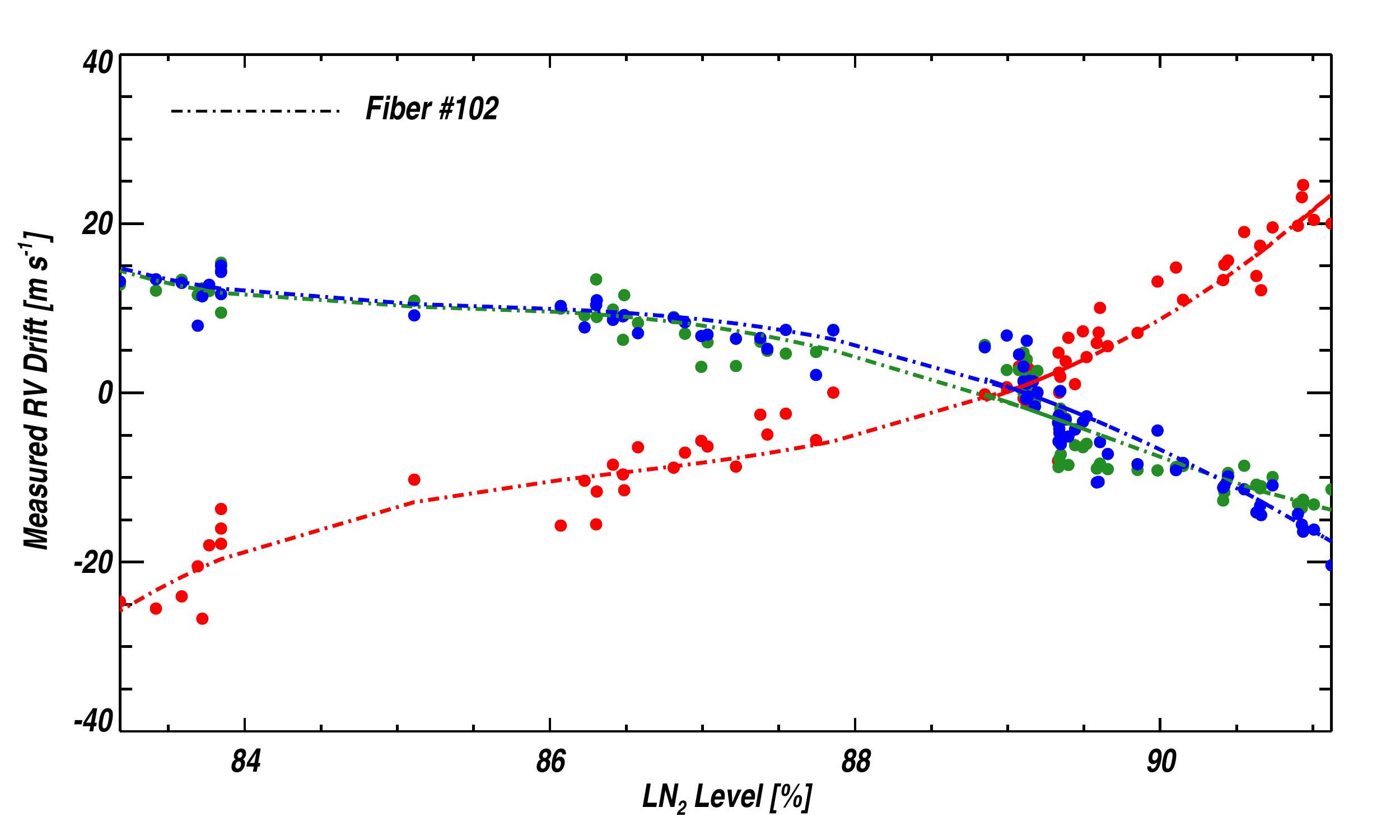}
\includegraphics[width=3.5in]{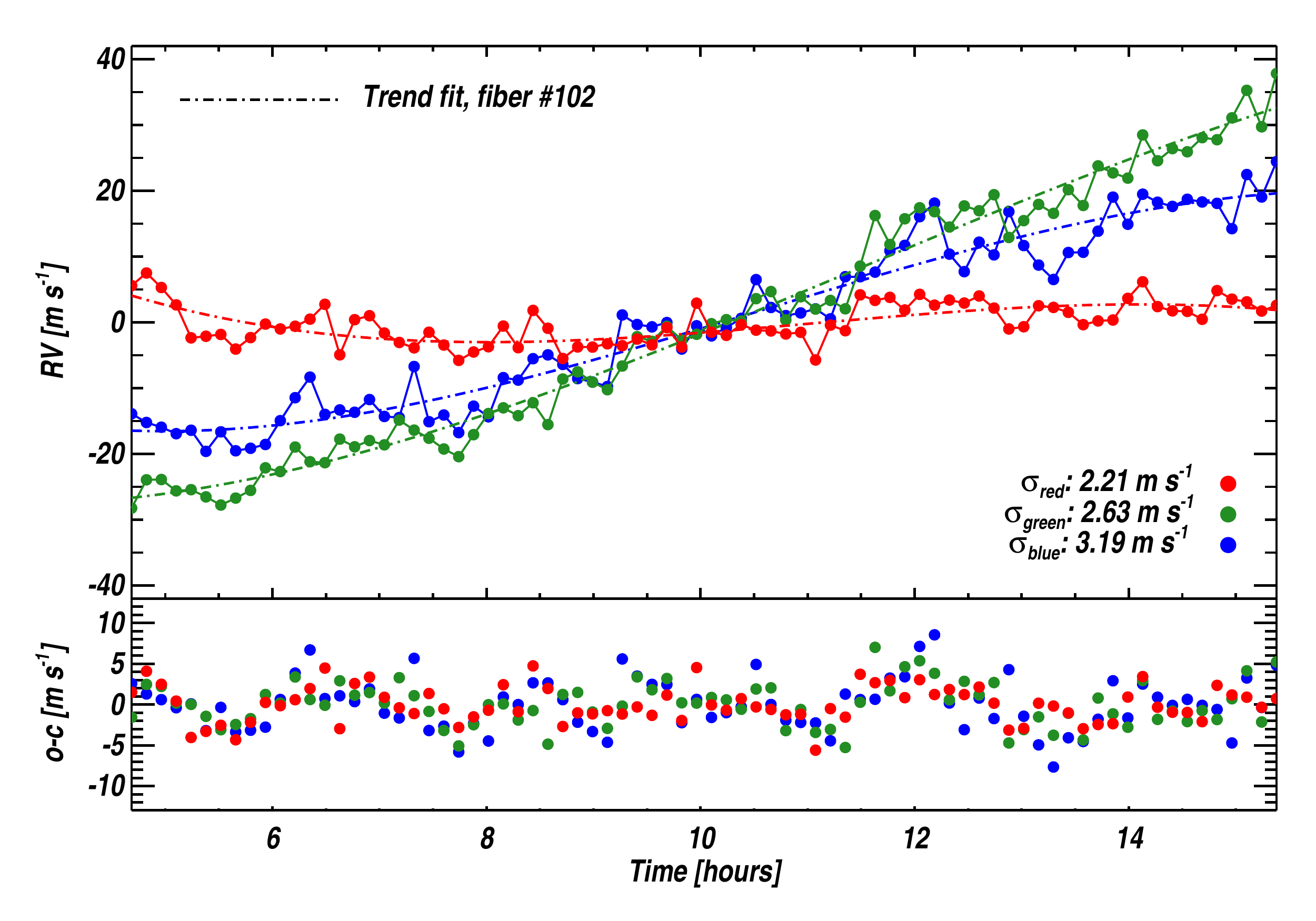}
\caption{Top: Observed spectrograph drift as function of LN$_2$ coolant level. Note that measured drift is significantly different for each detector. Additionally, the correlation varies spatially across any given detector, implying the drifts are not related to the FFP.  Bottom: Radial velocity drift of single APOGEE fiber over 12 hour interval for each of the three chips in the detector mosaic. The RMS scatter for each fiber is roughly 2-3 m s$^{-1}$  after removal of a low-order polynomial fit to the data series. }
\label{fig:apogee_ffp_rv}
\end{center}
\end{figure}

A distinct correlation exists between observed spectral drifts in the APOGEE focal plane and fill level of the liquid Nitrogen (LN$_2$)\footnote{A capacitance-based liquid level sensor and controller from American Magnetics\footnotemark[0] are used to measure the cryogen level in real-time.} used to cool the instrument. The LN$_2$ tank is attached underneath the optical bench of the instrument and is topped-off each morning by an automated fill system to maintain the 80 K cryostat operating temperature. As the LN$_2$ boils off throughout the day, the change in weight load in the storage tank results in a small but measurable flexing of the optical bench. This flexing affects the optical path enough to induce shifts in the focal plane in both spatial and spectral dimensions. This effect had been noted by the APOGEE team previously and is seen clearly in the FFP data. The magnitude of this correlated RV signal matches previous measurements using a combination of atmospheric airglow lines to track the spectral drift as a function of coolant fill-level. This systematic drift is corrected by removing a low-order polynomial trend in the RV time series (see Figure~\ref{fig:apogee_ffp_rv}). While this trend removal is likely accounting for other systematics beyond the LN$_2$-fill flexure of the optical bench, such as detector temperature changes and pressure variations within the LN$_2$ tank, we do not believe the FFP contributes to this signal as the trend varies significantly between each detector and spatially across different fibers. 

\begin{figure*}
\begin{center}
\includegraphics[width=3.15in]{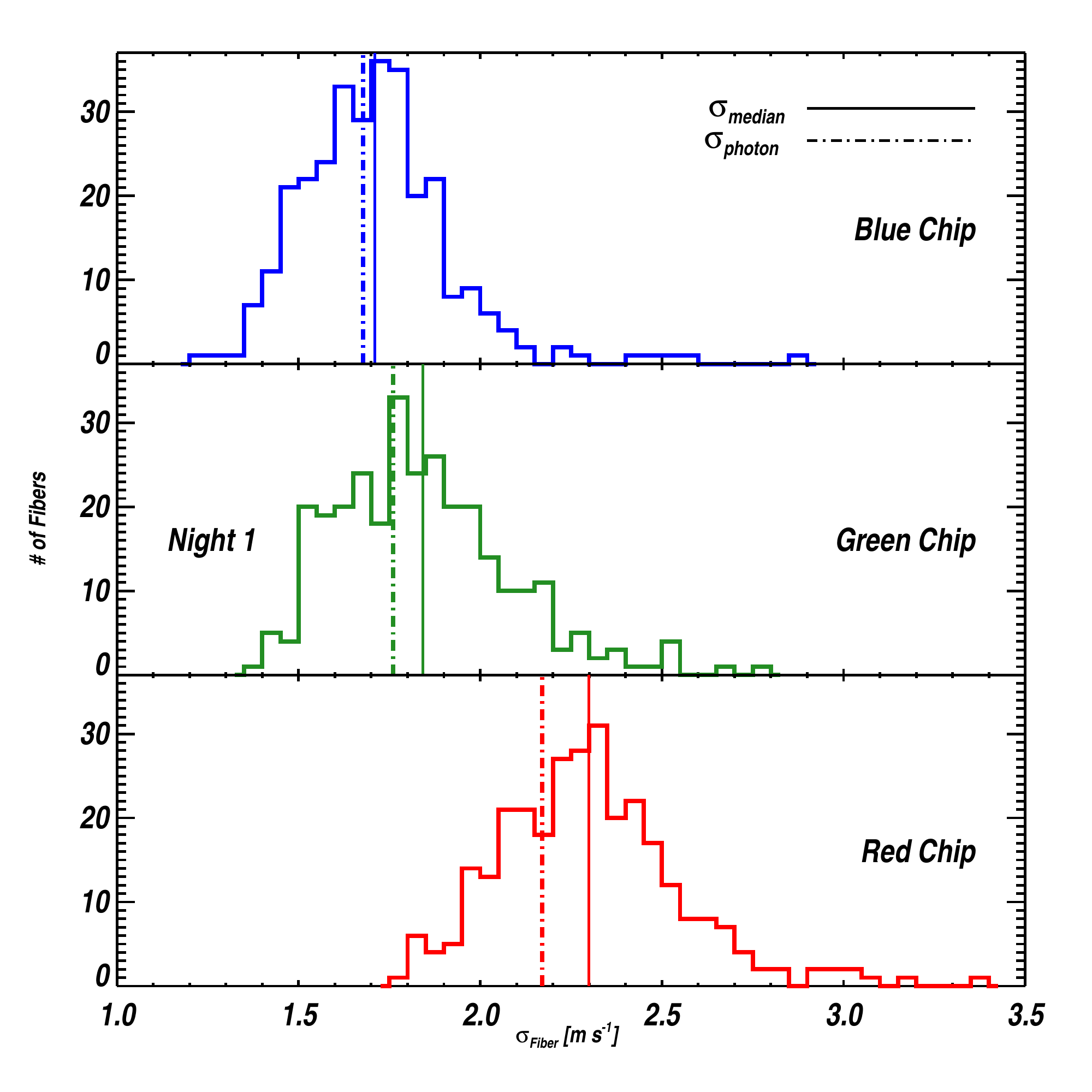}
\includegraphics[width=3.15in]{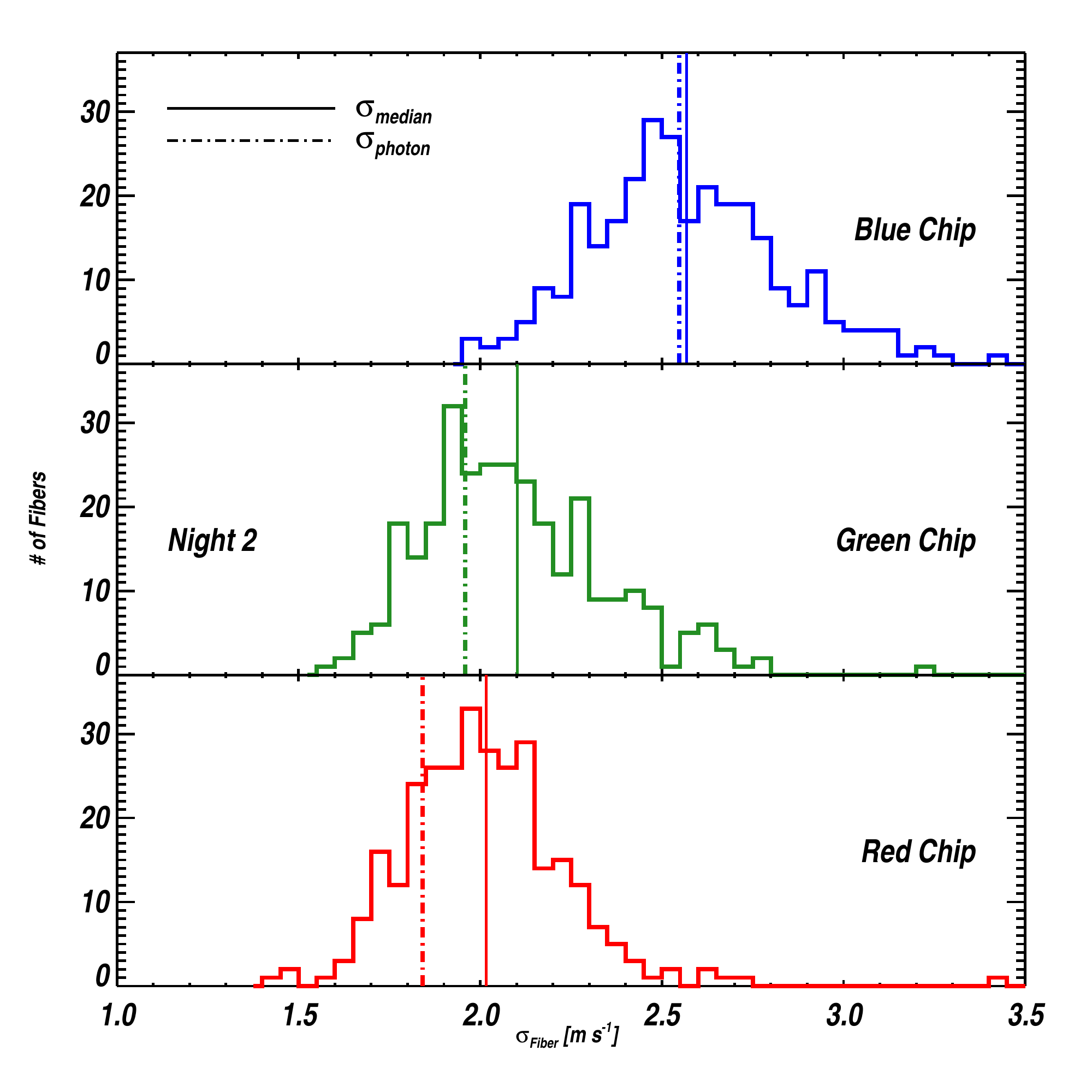}
\caption{Residual velocity scatter for all 300 APOGEE fibers illuminated with the FFP for both nights of observations after removal of  low-order spectrograph drift  and pressure correlated RV signal. Over-plotted are the expected photon-limited precisions (dashed line) as calculated using Equation~\ref{eq:qfact}. The FFP intrinsically seems to add very little short term scatter to the overall RV precision. }
\label{fig:photon_hist}
\end{center}
\end{figure*}

 A residual scatter of 1 - 3 m s$^{-1}$ is measured in each fiber after the spectrograph drift is removed for both nights of data. 
This precision is close to the expected photon-limited velocity precision (see Figure~\ref{fig:photon_hist}), though the residuals for the second night still show signs of systematic variations. The fiber-averaged residuals over the first night of data show a residual scatter of 80 cm s$^{-1}$ over a 12 hour period with no significant remaining structure.

\subsection{Cryostat Pressure - RV Correlation} 
\begin{figure}
\begin{center}
\includegraphics[width=3.5in]{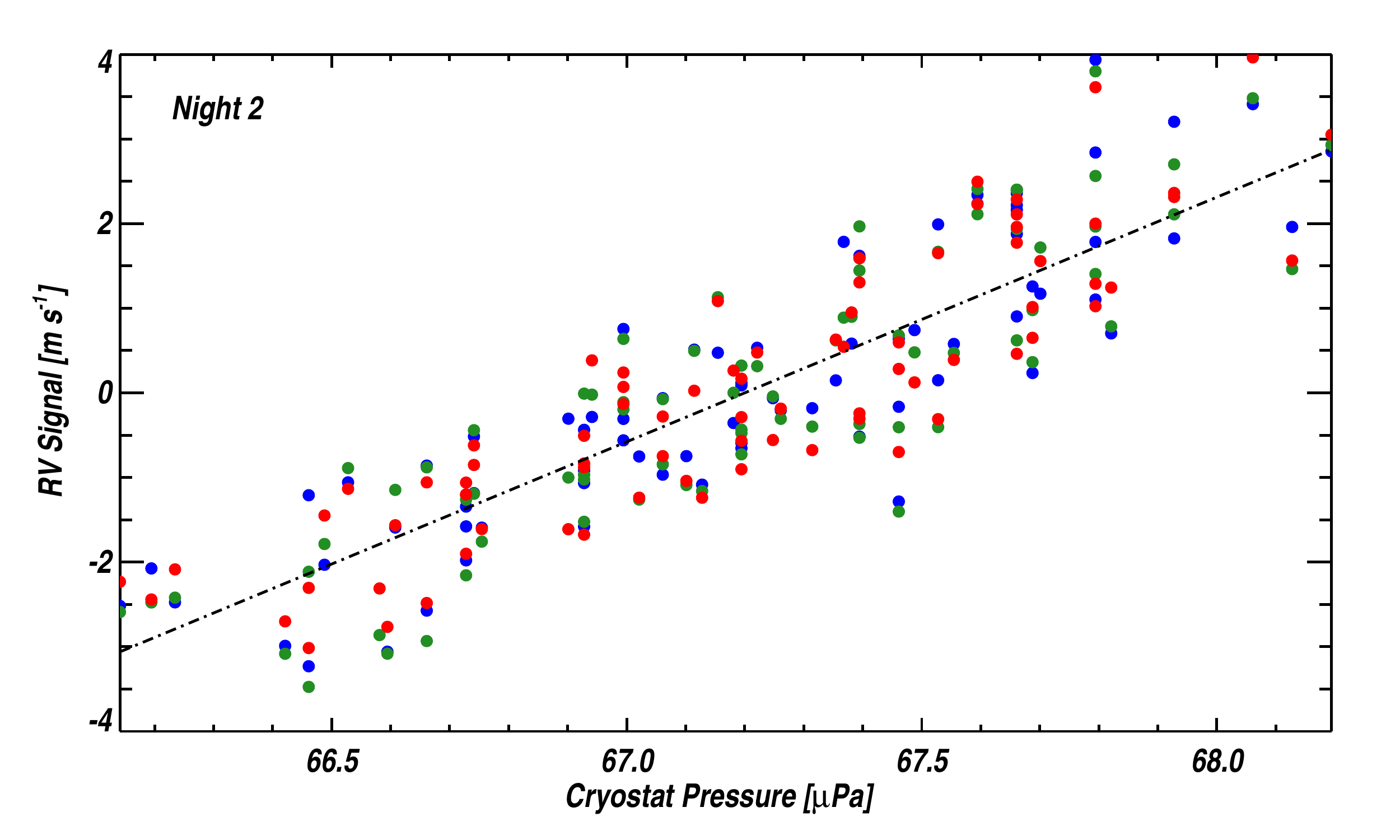}
\caption{Correlation between internal cryostat pressure and measured RV signal in FFP data during second night of observations for each APOGEE detector. A clear correlation is present at the several m s$^{-1}$ level. }
\label{fig:apogee_ffp_pressure}
\end{center}
\end{figure}

\begin{figure*}
\begin{center}
\includegraphics[width=3.5in]{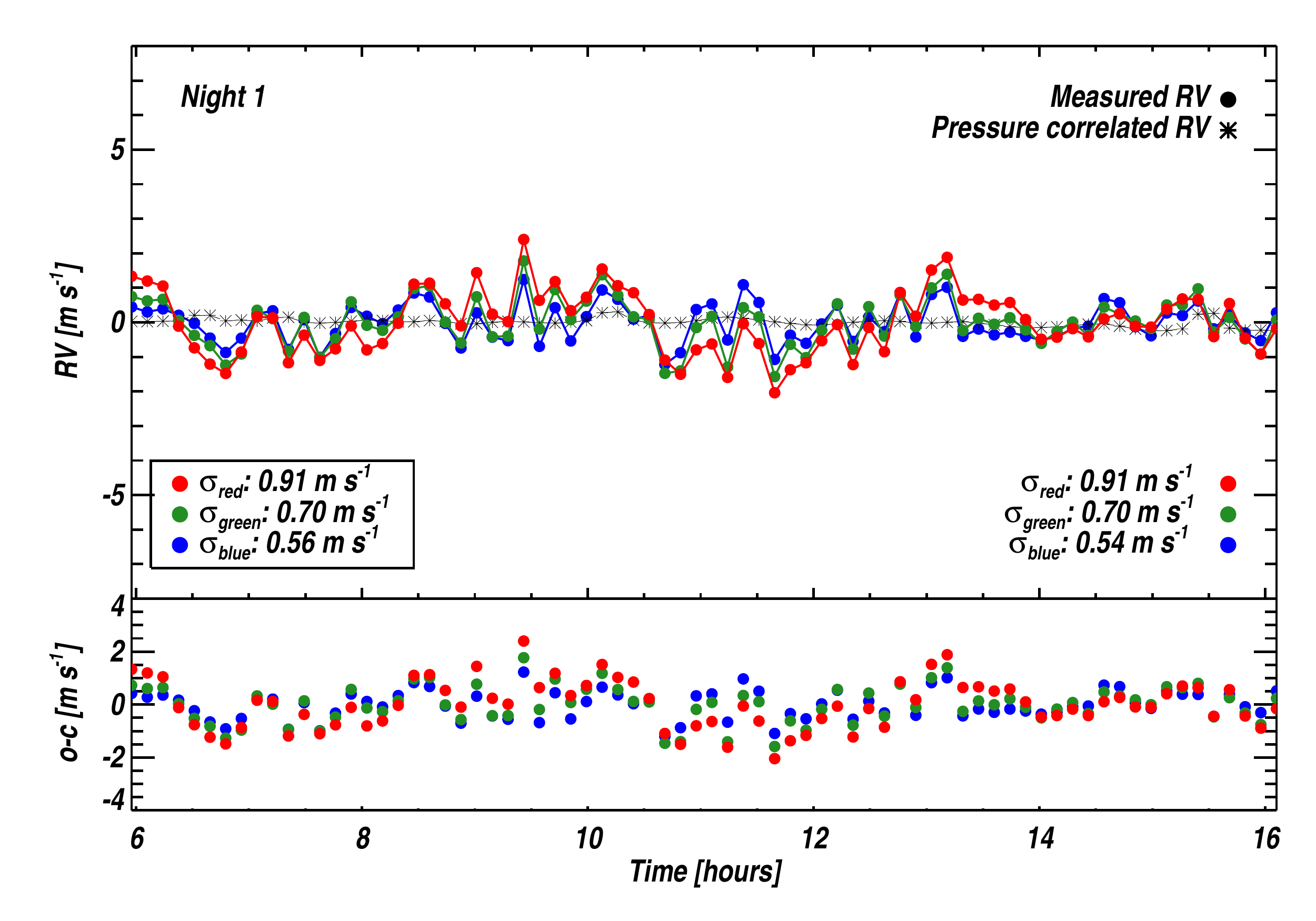}
\includegraphics[width=3.5in]{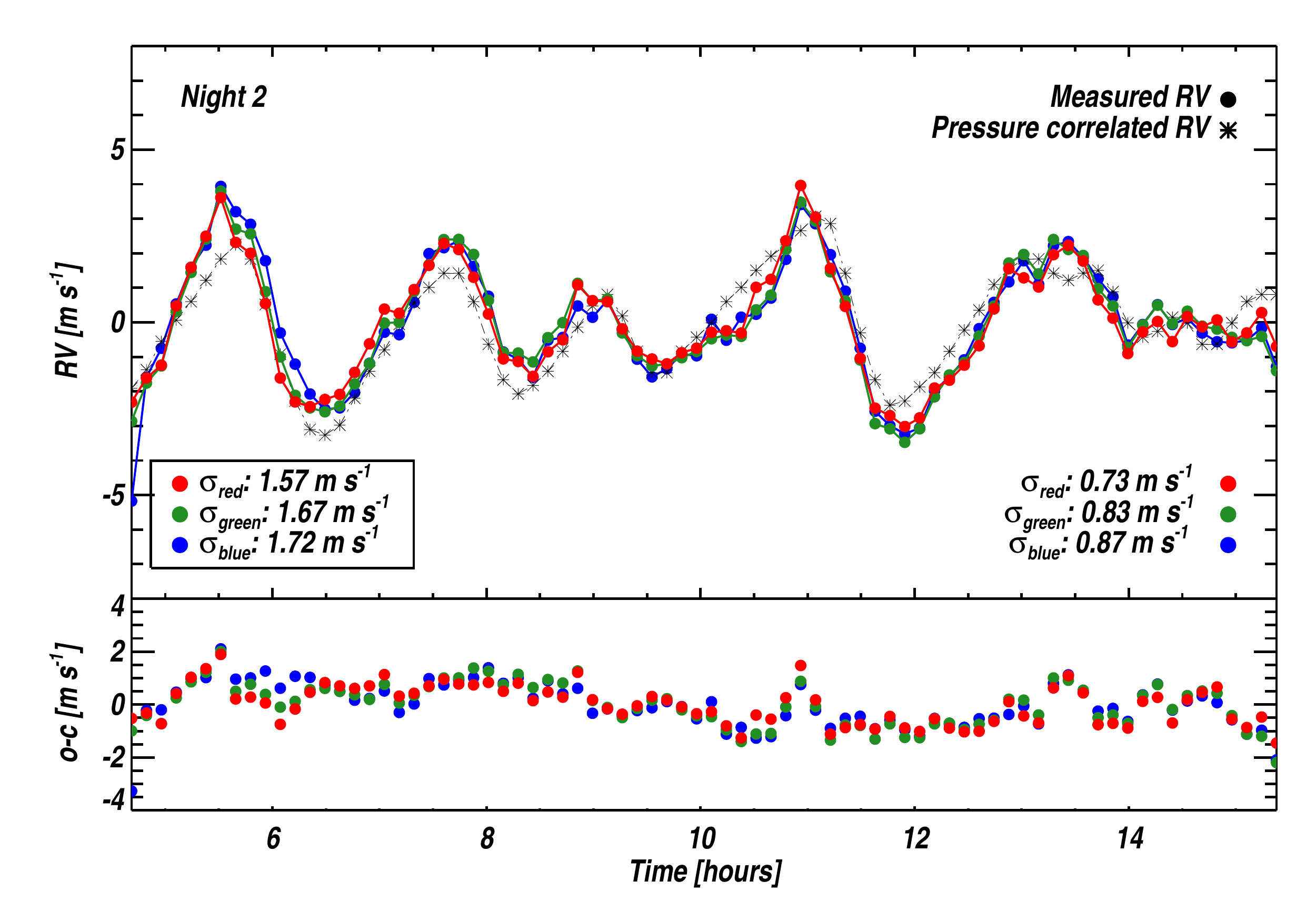}

\caption{Averaged residuals for all 300 APOGEE fibers with low-order  trend removed for both nights of data. Residual values prior to removal of pressure-correlated drift are shown in the boxed plot labels. The first night shows no significant pressure variations. The residuals for both nights reduce to roughly 80 cm s$^{-1}$ or better for all three-detectors once the pressure signal is subtracted.}
\label{fig:apogee_ffp_rv_final}
\end{center}
\end{figure*}

The remaining velocity variations observed on the second night correlate strongly with measured pressure variations inside the APOGEE cryostat\footnote{Two pressure probes are used to monitor the internal vacuum pressure of the cryostat, an MKS\footnotemark[0] 972 DualMag$^\mathrm{TM}$ Cold-Cathode MicroPirani$^\mathrm{TM}$ and a Pfeiffer\footnotemark[0] Full Range Pirani Cold Cathode gauge. Both gauges agree to within respective errors.}. The correlation becomes more obvious when averaging the residuals for all 300 fibers (see Figure~\ref{fig:apogee_ffp_pressure}.) During the second night of data collection, the pressure inside the cryostat varied periodically during the observation period. These variations were very small, on the scale of 10 $\mu$Pa (75 nano-Torr), but correlate strongly with the observed spectral drifts in the FFP data. The residual scatter after removing this pressure-correlated RV signal is $\approx$80 cm s$^{-1}$ for each night, approaching the expected photon-noise limit of 20 cm s$^{-1}$ (see Figure~\ref{fig:apogee_ffp_rv_final}.)

 The pressure variations inside the cryostat are too small in amplitude to change the refractive index (and by extension RVs) of the residual gas inside the cryostat at any measurable level and do not correlate with mountain temperature or pressure fluctuations, implying that the internal variations must be a proxy for some other environmental parameter affecting the optical bench. We resist the temptation to explain this effect here, as the goal of this study is to determine the intrinsic stability of the FFP.
 
We do not believe there is any significant pressure sensitivity in the FFP as the central fiber section is epoxied into a ferrule. The ferrule and epoxy encasement effectively shield the short section of SMF that is susceptible to any pressure-induced strain. Furthermore, we see no correlation between the measured shifts in the FFP spectra and recorded mountain pressure.

\section{Fourier Transform Spectometer Scans}
Unlike a frequency-locked LFC, the interferometric nature of the FFP does not by itself provide an absolute wavelength reference. A stabilized laser, atomic emission lamp, or molecular absorption cell reference is needed for absolute calibration of the etalon. Absolute wavelength measurements of astronomical Doppler calibration sources are usually done with stabilized high resolution spectrographs (e.g. Th-Ar measurements by \cite{2006SPIE.6269E..23L}).

To derive an absolute wavelength solution for the FFP, high resolution scans were taken using the 2.0 m Fourier Transform Spectrometer (FTS) at the National Institute of Standards and Technology (NIST). The aim of the FTS scans was to accurately measure the line centers of the FFP spectrum and calibrate against a NIST standard reference molecular gas cell. Ideally, the measured wavelength solution could then be applied to any future FFP spectra. C$_2$H$_2$ and HCN cells were used to derive precise wavenumber corrections for the FTS before and after each scan.

Initial spectra of the FFP output using the supercontinuum illumination source were unexpectedly noisy, even after combining a large number of coadded scans (S/N $<$ 50). The high noise level is attributed to both the quasi-stable pulsation rate of the Q-switching laser used in the supercontinuum generation, and the very bright emission line at the 1064 nm pumping wavelength. Using an oscilloscope, we observed the repetition rate of the laser to vary by between 5 \% and 10 \% around the 24 kHz laser repetition frequency. Since the NIST 2-m FTS takes data at equal time intervals, fluctuations in the frequency of the supercontinuum source add significant noise to the interferogram and resulting spectrum. Addition of a 1100 nm long-pass filter reduced the measured noise level due to the 1064 nm spike, but still resulted in insufficient signal-to-noise in the final spectra for precise characterization of line centers.

Several other continuum sources were tested, including higher power supercontinuum sources with more stable repetition rates and NIR superluminescent diodes (SLEDs). Scans taken using these sources yielded significantly higher signal-to-noise spectra, though we did measure large drifts ($>$100 m s$^{-1}$) between spectra taken with different illumination sources. We tentatively attribute this drift sensitivity to both the difference in polarization output of each continuum source and the change in input fiber configuration prior to the interferometer cavity between scans. 

\section{Polarization Sensitivity}

\begin{figure}[!h]
\begin{center}
\includegraphics[width=3.4in]{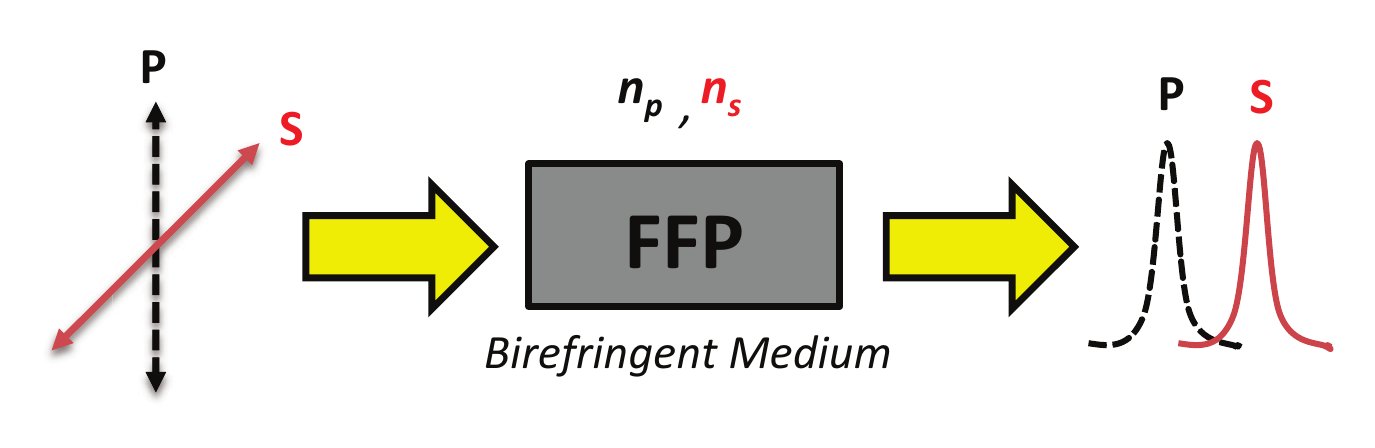}
\includegraphics[width=3.4in]{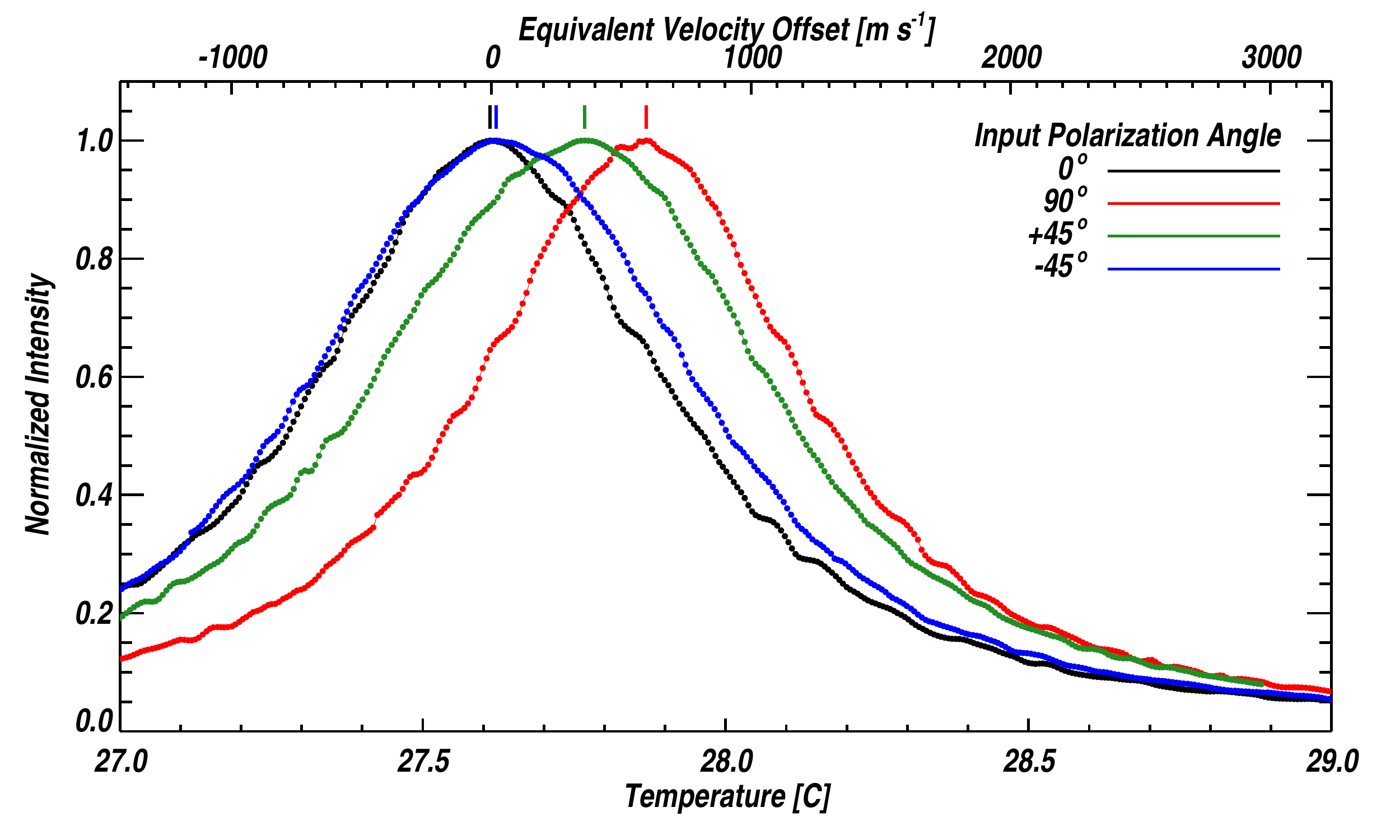}
\caption{Top: Diagram outlining FFP polarization sensitivity. The strong birefringence of the SMF-28 fiber dominates the line locations. The two refractive indices produce two separated Airy peaks. Bottom: Intensity-temperature scans of the SMF-28 FFP. A linear polarizer was inserted prior and rotated between four different positions (vertical, horizontal, +45 degrees and -45 degrees). A precise temperature ramp was used to map the Airy peaks for each input polarization using a narrow-band (10 MHz) 1550 nm diode laser as the illumination source. The output intensity was measured using a photodiode while the cavity temperature was steadily increased.}
\label{fig:ffp_pol}
\end{center}

\end{figure}

Single-mode fibers, like most optical fibers, can be birefringent. For any birefringent material, there is a different refractive index for the $S$ and $P$ polarizations ($n_\mathrm{S}$ and $n_\mathrm{P}$). In the case of a Fabry-Perot, this leads to an output spectrum that is dependent on the incident polarization state (see Figure~\ref{fig:ffp_pol}). Imperfections, stresses, or temperature variations in the fiber can dramatically alter the polarization state of any input source.  A polarized illumination source can have its original polarization state altered to any random axis by propagating through a short section of fiber. Our initial device configuration did not precisely control the polarization state of light incident on the cavity, which can lead to the large spectral drifts between different input fiber configurations as seen in the FTS data. Both the location and width of the Airy peaks change significantly between polarization states. 

Laboratory measurements of this effect are shown in Figure~\ref{fig:ffp_pol}. A narrowband, fixed-wavelength 1550 nm laser is used to illuminate the FFP as the temperature of the cavity was steadily increased through a single Airy peak. This temperature scanning technique allows for a very high resolution spectrum of the intrinsic interferometer profile. A linear polarizer was inserted between the supercontinuum illumination source and the FFP input and rotated between scans.

\begin{figure}
\begin{center}
\includegraphics[width=3.4in]{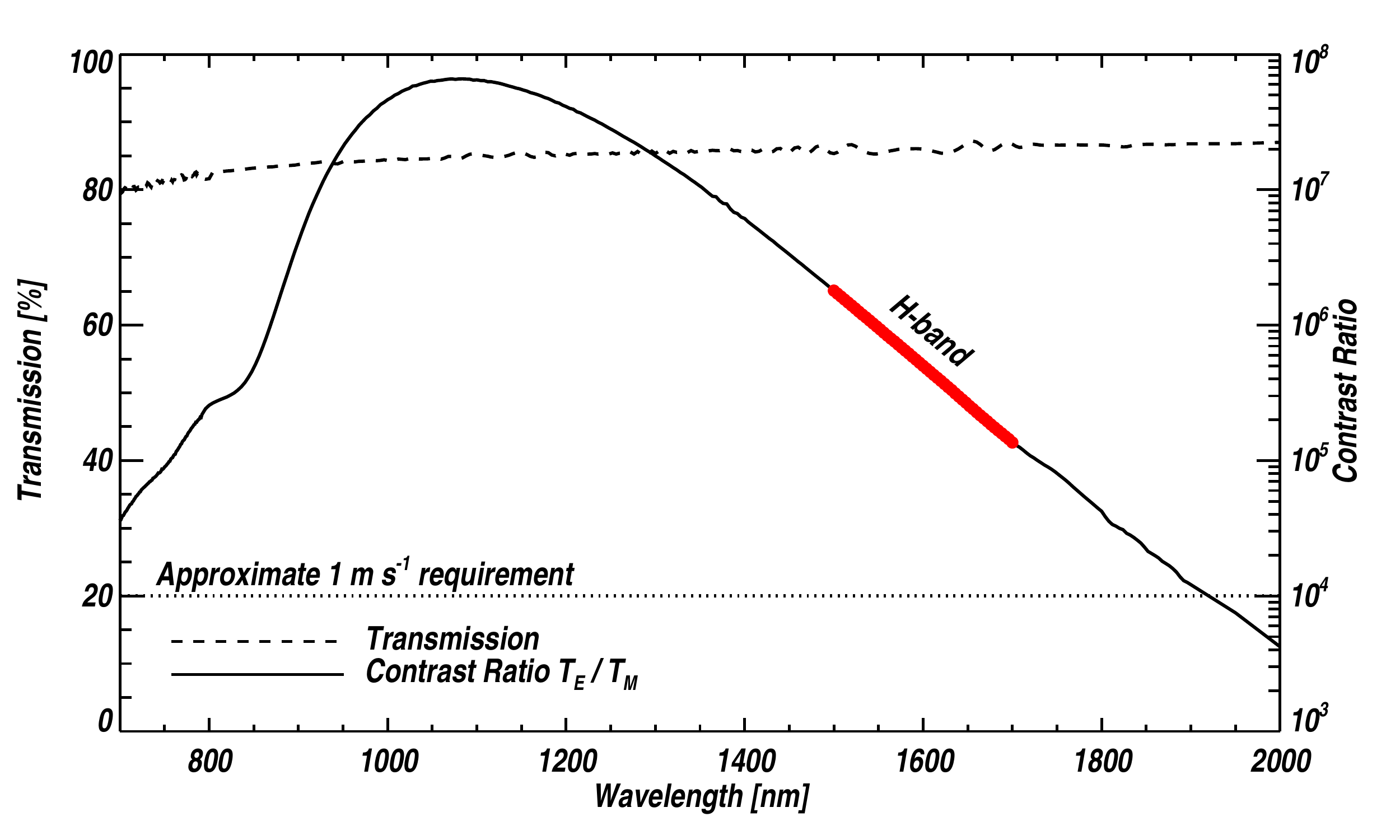}
\caption{Transmission and contrast ratio of NIR high contrast linear polarizer used to stabilize polarization state incident on FFP cavity. The polarizer provides high contrast ratios in all NIR bands, and $>$10$^5$ in the H-band.} 
\label{fig:lpnir_pol}
\end{center}
\end{figure}

The solution to this polarization dependence is to ensure a single, stable polarization state is always incident on the interferometer cavity input. For the FFP described here, a contrast ratio of roughly $10^{4}$ between orthogonal polarization states propagating through the cavity is required to maintain output spectral stability at the 10's cm s$^{-1}$ level based on our laboratory measurements. Polarization maintaining  (PM) fibers were initially considered as viable options to control the input polarization, but do not provide sufficient extinction ratios and are intrinsically highly temperature sensitive.
Our solution to this birefringence issue is to insert a commercial ThorLabs\footnotemark[0] high-contrast linear polarizer\footnote{Thorlabs\footnotemark[0] LPNIR polarizer} prior to the FFP input.  Figure~\ref{fig:lpnir_pol} shows the typical contrast curve of the polarizer used.

\begin{figure}
\begin{center}
\includegraphics[width=3.4in]{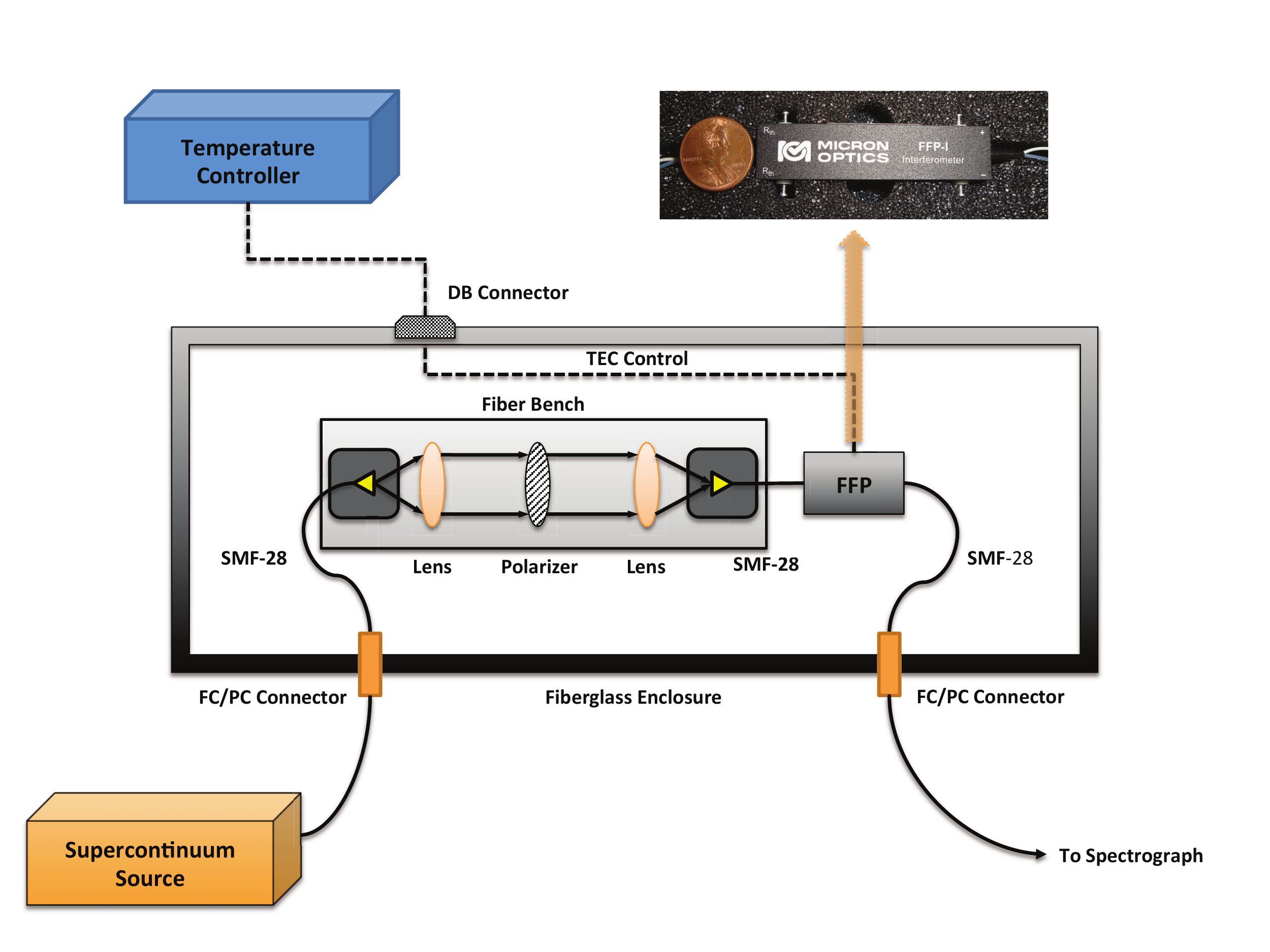}
\includegraphics[width=3.4in]{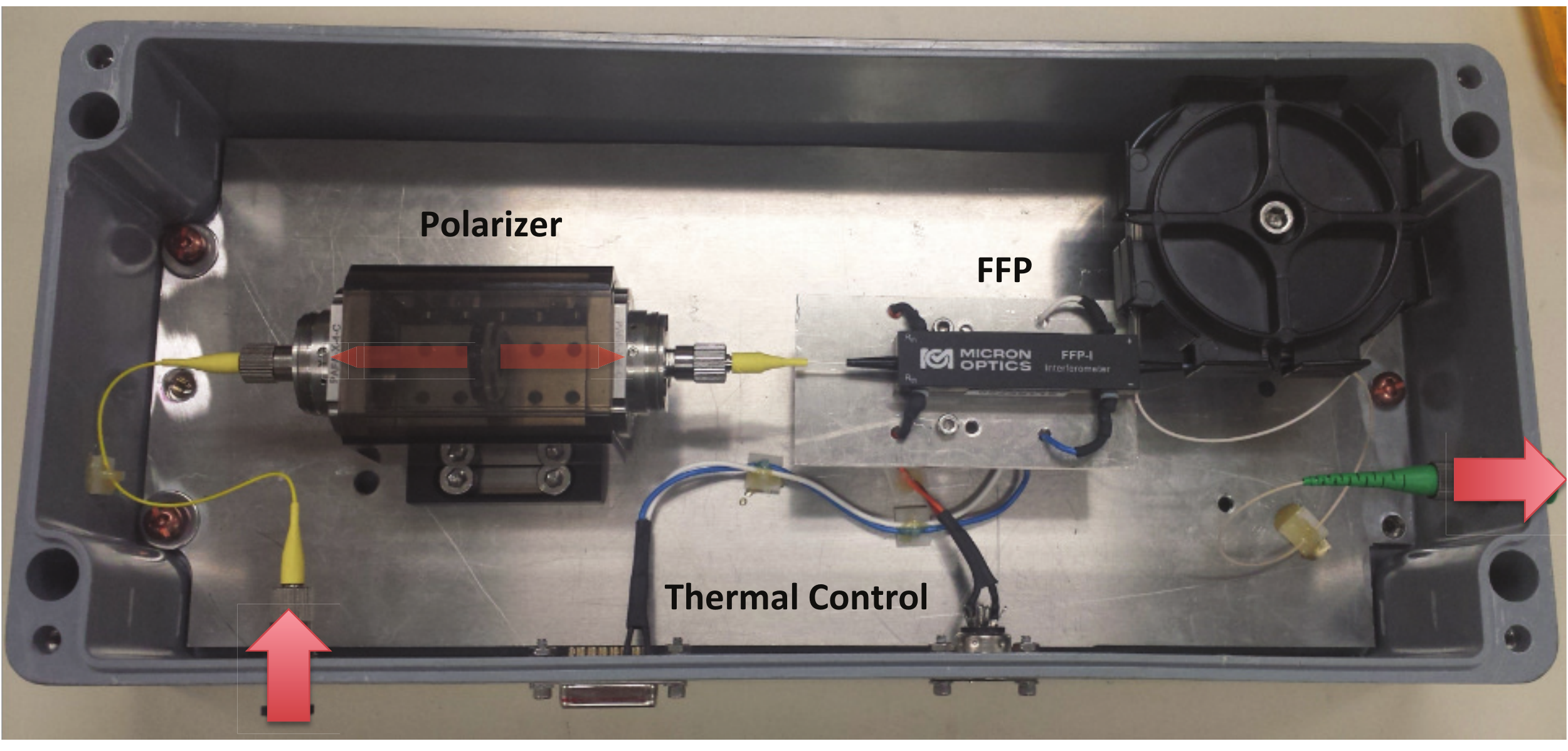}

\caption{Top: Schematic of FFP configuration with linear polarizer. Input light is collimated on a small fiber bench, passed through a high-contrast linear polarizer, and focused onto a short section of fiber. This setup ensures a single polarization state is incident on the FFP input. Bottom: Image of FFP system with linear polarizer.} 
\label{fig:ffp_encl}
\end{center}
\end{figure}

Light from the supercontinuum source is first refractively collimated on a small fiber bench and then passed through the high-contrast polarizer. The collimated beam is then refocused onto the input fiber of the FFP.
The input fiber of the FFP was trimmed from 2 m to approximately 5 cm to reduce the amount of birefringent material prior to the interferometer. The modified FFP is shown in Figure~\ref{fig:ffp_encl}. This allows for a much more rigid input configuration and significantly reduces the device's sensitivity to source polarization and fiber birefringence.

The polarization modifications were not completed prior to the APOGEE observing period, though we do not believe the fiber birefringence was an issue over the course of our experiments as the supercontinuum source is unpolarized and the input fiber configuration was not altered over a given night of data collection.

\section{Summary}
We present a NIR H-band fiber Fabry-Perot interferometer as a precise spectrograph calibration device. This device is a commercially available product we are developing as a prototype for a similar calibration source being developed for the HPF instrument. An FFP is a unique alternative to traditional wavelength calibration systems, providing a dense set of stable spectral features across a wide wavelength range. A stabilized fiber etalon can yield high relative precisions at relatively low cost.

Precise temperature control is essential for these devices, as the spectral output is sensitive to significant shifts even at moderately stable temperature control. With a benchtop controller, we are able to achieve a thermal control precision of 100 $\mu$K, corresponding to roughly 22 cm s$^{-1}$ velocity stability. Tests on the SDSS-III APOGEE spectrograph showed the FFP was able to track instrument drift at precisions close to the expected photon-limit of the instrument. An individual APOGEE fiber was tracked to better than $\approx$3 m s$^{-1}$. This represents nearly two orders of magnitude improvement in measurement sensitivity for the APOGEE instrument. When averaging over all fibers, our measurement precision improves to 80 cm s$^{-1}$ over 12 hours.

Stress-induced birefringence of the SMF used in the FFP leads to a strong polarization dependence in the output spectrum. This effect was directly observed in the laboratory experiments using a polarized 1550 nm laser in data taken at the NIST 2.0 m FTS, though we do not believe this was an issue in the APOGEE data discussed here. We have significantly reduced this sensitivity by adding a high-contrast polarizer at the FFP input and shortening the amount of polarization-altering SMF-28 fiber prior to the interferometer cavity.

The high RV precisions achieved here validate the design attributes of the APOGEE instrument as ideal fundamental traits for a highly stable Doppler velocimeter. Many of these ideas will be used on the HPF instrument to achieve better stability and overall velocity precision. These results also validate the fundamental soundness of the cryogenic enclosure design being adopted for precision RV instruments like HPF.

\section{Future Work}
FFP designs that are optimized for the z/Y/J NIR bands are being explored for the HPF instrument. The difficulty lies in manufacturing a FFP that is single-mode over the entire HPF bandpass, has high finesse, and is insensitive to both ambient temperature fluctuations and illumination source polarizations. The first point is an issue of fiber selection and we are actively exploring different fiber options from a number of vendors that are single-mode across the HPF wavelength range. To achieve the optimum finesse and mirror bandwidth, several dielectric mirror configurations are being considered. The ideal mirrors would have a high, flat reflectivity response between 0.8-1.3 $\mu$m. To desensitize the cavity from input polarization, the next iteration of the device will have a high contrast thin film polarizer at the input of the interferometer cavity.  Placing the polarizing element in direct contact with the interferometer section ensures the polarization will be decoupled from the input fiber birefringence and illumination source polarization. This placement does increase the risk of thermal fluctuations due to increased absorption in the polarizing film, though with good thermal control this effect will be minimal. Additionally, we have chosen to use a high-efficiency wire-grid reflective polarizer, rather than an absorptive polarizer, to mitigate the risk of significant power absorption.

\acknowledgments{This work was partially supported by funding from the Center for Exoplanets and Habitable Worlds. The Center for Exoplanets and Habitable Worlds is supported by the Pennsylvania State University, the Eberly College of Science, and the Pennsylvania Space Grant Consortium. We acknowledge support from NSF grants AST 1006676, AST 1126413, AST 1310885, and the NASA Astrobiology Institute (NNA09DA76A) in our pursuit of precision radial velocities in the NIR. SPH acknowledges support from the Penn State Bunton-Waller, and Braddock/Roberts fellowship programs and the Sigma Xi Grant-in-Aid program. This research was performed while SLR held a National Research Council Research Associateship Award at NIST. 

This work was based on observations with the SDSS 2.5-meter telescope. Funding for SDSS-III has been provided by the Alfred P. Sloan Foundation, the Participating Institutions, the National Science Foundation, and the U.S. Department of Energy office of Science. The SDSS-III web site is \url{http://www.sdss3.org/}. SDSS-III is managed by the Astrophysical Research Consortium for the Participating Institutions of the SDSS-III Collaboration including the University of Arizona, the Brazilian Participation Group, Brookhaven National Laboratory, University of Cambridge, Carnegie Mellon University, University of Florida, the French Participation Group, the German Participation Group, Harvard University, the Instituto de Astrofi\'sica de Canarias, the Michigan State/Notre Dame/JINA Participation Group, Johns Hopkins University, Lawrence Berkeley National Laboratory, Max Planck Institute for Astrophysics, Max Planck Institute for Extraterrestrial Physics, New Mexico State University, New York University, Ohio State University, Pennsylvania State University, University of Portsmouth, Princeton University, the Spanish Participation Group, University of Tokyo, University of Utah, Vanderbilt University, University of Virginia, University of Washington, and Yale University. }

\bibliographystyle{apj}
\bibliography{ms}

\end{document}